\documentclass[graybox]{svmult}

\usepackage{type1cm}        
\usepackage{tikz}
\usepackage{hyperref}                           
\usepackage{makeidx}         
\usepackage{graphicx}        
\usepackage{multicol}        
\usepackage[bottom]{footmisc}
\usepackage{color}

\usepackage{newtxtext}       %
\usepackage[varvw]{newtxmath}       

\makeindex             

\begin{document}
	
	\title*{Social human collective decision-making and its applications with brain network models}
	\author{Thoa Thieu and Roderick Melnik}
	\institute{Thoa Thieu \at MS2Discovery Interdisciplinary Research Institute, Wilfrid Laurier University, Waterloo, Ontario, Canada, \email{tthieu@wlu.ca}
		\and Roderick Melnik \at MS2Discovery Interdisciplinary Research Institute, Wilfrid Laurier University, Waterloo, Ontario, Canada  \& BCAM - Basque Center for Applied Mathematics, Bilbao, Spain \email{rmelnik@wlu.ca}}
	%
	%
	\maketitle
	
	\abstract*{}
	
	\abstract{	A better understanding of social human dynamics would be a powerful tool to improve nearly any computational endeavour that involves human interactions. This includes intelligent environments featuring, for instance, efficient illumination systems, smart evacuation signalling systems, intelligent transportation systems, crowd control, or disaster response. Moreover, given that the human population has significantly grown up in number and spread across the planet, the capacity to predict social human behaviours will help to demonstrate special behavioural forms observed when masses of people gather together and make crowds. Additionally, human crowd dynamics are characterized by complex psychological and sociobiological behaviour. The contributions of psychological factors need to be accounted for to obtain more reliable models. Many models have been proposed to describe the social human group dynamics in different scenarios. However, due to the complexity of such systems amplified by the above factors, social human decision-making with multiple choices has not been
		fully scrutinized. In this chapter, we consider probabilistic drift-diffusion models and Bayesian inference frameworks to address this issue, assisting better social human decision-making. We provide details of the models, as well as representative numerical examples, and discuss the decision-making process with a representative example of the escape route decision-making phenomena by further developing the drift-diffusion models and Bayesian inference frameworks. In the latter context, we also give a review of recent developments in human collective decision-making and its applications with brain network models. Furthermore, we provide illustrative numerical examples to discuss the role of neuromodulation, reinforcement learning in decision-making processes. Finally, we call attention to existing challenges, open problems, and promising approaches in studying social dynamics and collective human decision-making, including those arising from nonequilibrium considerations of the associated processes.}
	
	
	\section{Introduction}\label{intro}
	
	In recent years, many models have been proposed to describe the decision-making of human crowd dynamics in different scenarios. The fields of human crowd dynamics consist largely of works not only 
	in mathematics, scientific computing, engineering, and
	physics but also in social psychology. The detailed
	behavior of human crowds is already complicated. Many physiological
	and sociobiological processes act with the physical feedback mechanism effects caused by the surrounding environments, which are still largely unknown. To develop reliable human crowd dynamic models, the contributions of psychological factors need to be accounted for. There is overwhelming evidence in support of this. For example, panic in crowd dynamics is often caused by attempting escape of pedestrians from threats in situations of a perceived struggle for survival or eventually ending up in trampling or crushing. Hence, decision-making with many social learning processes, including opinion dynamics, is essential in studying human group movements. In general, models of human decision-making from the brain research point of view are important in studying cognitive psychology and have been successfully used to fit experimental data and relate them to neurophysiological mechanisms in the brain. This research direction raises a relevant question what is brain neuroscience revealing about social decision-making? Recall now that one of the most important models for binary decision-making is the drift-diffusion
	model (DDM) \cite{Fudenberg2020}. On the other hand, DDM includes a diffusion process that models the accumulation of perceived evidence and yields the decisions upon reaching specified thresholds \cite{Vellmer2020}. On the other hand, DDMs may
	bridge the gap between experiments of decision-making and neurobiologically
	motivated models that describe how the decision-making process is implemented in the brain. In such models, evidence,
	in the form of sensory information, enters competing neural
	networks, mimicking the work of real brain networks. For instance, information encoded by (approximately
	Poissonian) spike trains of neurons are accumulated by a neural
	population. This accumulation can be approximated by an
	Ornstein–Uhlenbeck process \cite{Smith2015}. However, the next level of complexity comes from the fact that the characteristics of human group dynamics are affected by the decisions of each individual in the group. In making such decisions, and in particular when the decision context involves certain degrees of uncertainty, humans tend to utilize all possible sources of information notably social information. Humans often resort to the
	decisions made by others as additional sources of information to improve
	their decision-making. This could potentially link to opinion and/or learning dynamics among human groups \cite{Mavrodiev2013,Mavrodiev2021}. When this process is followed by a tendency to
	‘imitate’ the majority's decision, it could lead to unplanned
	coordination of the actions, often referred to as ‘herd’ behaviour \cite{Haghani2019}. The Bayesian model provides an appropriate tool to explain how the brain extracts information from noisy input as typically presented in perceptual decision-making tasks. Additionally, it is now known that the DDM has a relationship with such functional Bayesian models \cite{Gold2007,Bogacz2006,Bitzer2014,Fard2017}.

	%
	
	
	We also know that the brain infers our spatial orientation and properties of objects in the world from ambiguous and noisy sensory cues. Moreover, the recognition of self-motion in the presence of stationary, as well as independently moving objects, offers many challenging inference problems. This is due to the fact that the image motion of an
object could be attributed to the movement of the object, self-motion,
or some combination of the two \cite{Dokka2019}.  On the other hand, we note that human perception (the process whereby sensory stimulation is translated into organized experience, e.g., vestibular signals) play a critical role in dissociating self-motion
from object motion \cite{Dokka2015}. Hence, motivated by \cite{Dokka2015,Dokka2019,Noel2022}, one of our representative examples of the developed theory based on DDM and Bayesian approach, will be the decision-making process with application to risky escape route decision phenomena, along with the analysis of other important aspects of this process. 

The rest of this chapter is organized as follows. In Section \ref{sec:DDMBayesian}, we consider probabilistic DDM and Bayesian inference frameworks for social human decision-making, where we provide details of the models, their brief survey, and theoretical foundations. Section \ref{sec:examples} is devoted to representative numerical examples, where we discuss the decision-making process in the risky escape route decision phenomena by considering the drift-diffusion models and Bayesian inference frameworks. Section \ref{sec:4collective} provides a review of recent results on human collective decision-making. Section \ref{sec:examplescollective} focuses on a numerical example and discusses the role of neuromodulation and reinforcement learning processes in studying decision-making processes. Section 6 is devoted to considering decision-making processes as nonequilibrium, similar to learning and knowledge creation, by focusing on human biosocial dynamics with complex psychological behaviour and nonequilibrium phenomena. Conclusions are given in Section \ref{sec:Conclusions}, where we highlight open problems and future research.

%
	\section{DDMs and Bayesian models for decision-making}\label{sec:DDMBayesian}
	
	\subsection{DDMs in probabilistic settings} \label{sec:DDM}

	While DDMs have been used in applications for a long time \cite{Ratcliff1978,Bitzer2014,Fudenberg2020}, their generalization to human decision-making problems is of recent origin.
The probabilistic DDM is described by sequential sampling with diffusion signals with Brownian motions. In particular, in probabilistic DDM, a decision is made by the following process: 
\begin{itemize}
	\item First, the decision maker accumulates evidence until the process hits either an upper or lower stopping
	boundary and then stops; 
	\item Second, the decision is made by choosing the alternative that corresponds to that boundary. 
\end{itemize}
Unlike many other applications, these problems require considerations of stochastic dynamics with boundary conditions \cite{Thieu2022coupled}.
Recently, the description of the decision-making processes in neuroscience and psychology has been proposed with probabilistic DDMs \cite{Fudenberg2020,Pekkanen2022}. For $t \in (0, \infty)$, the two main ingredients of our probabilistic DDM are the stochastic process $X(t)$ and a boundary function. Let us define the following system of drift-diffusion equations modelling the decision-making process as follows:

\begin{align}\label{DDM}
	d X(t) = \mu(X(t))dt  + \sigma(X) dB(t),
\end{align}
where $\mu \in C([0, \infty) \times \mathbb{R})$ represents the drift, while $\sigma\in (0,\infty)$ is diffusion coefficient. The term $B(t)$ denotes the standard Brownian motion. The initial condition for the system \eqref{DDM} are $X(0) = x_{0}$. For all $x_{0} \in [b(0), \tilde{b}(0)]$, we define the following hitting times of the boundaries $\alpha, \beta$:

\begin{align}
	\alpha = \inf\{t \geq 0: |X(t)| \leq b(t) \}, 
	\beta= \inf\{t \geq 0: |X(t)| \geq \tilde{b}(t) \}, \label{stoping_time} 
\end{align}
where $b, \tilde{b} \in C^1([0,\infty))$ satisfy $b \leq \tilde{b}$. The presentation \eqref{stoping_time} defines the first time when the absolute value of the process $X(t)$ hits the boundary $b$. In some cases, we can choose a specific boundary condition, e.g. reflecting boundary conditions or absorbing boundary conditions. In particular, the authors in \cite{Thieu2022coupled} have discussed the reflecting boundary conditions for a system of SDEs with applications in neuroscience. 

One of the reasons for a renewal of interest in drift-diffusion models in analyzing complex systems that includes human dynamics and behaviours is due to their statistical mechanics' foundations. They maintain a prominent role in a hierarchy of mathematical models derived from the Liouville equation for the evolution of the position-velocity probability density, representing the continuing interest of scientists in various areas of theory and applications \cite{Melnik2000,Melnik2000quasi,Melnik2000modelling,Estrada2002,Goddard2019}. Similar type equations have also been discussed in the realm of open systems preserving necessary thermodynamic consistency(e.g., \cite{Klein2022}). Moreover, under known simplifying assumptions, the derivation of the drift-diffusion model can further be rigorously justified, starting from a version of the Hilbert expansion. While other models within the mentioned hierarchy have also been used in crowd dynamics and related areas of active interacting particles (e.g., \cite{Degond2022,Jiang2022,Bellomo2022What,Bellomo2022towards} and references therein), we believe that for the field of interest here, it is essential to explore further the potential of probabilistic drift-diffusion models integrated with the Bayesian framework. Consolidating knowledge between mathematical modelling and cognitive science is necessary in this undertaking.
	
	In the context of collective human decision-making in particular and the collective behaviour of living species in general, the above model hierarchy and statistical mechanics play a critical role. One of the points of entry of these ideas into the description of collective dynamics has traditionally been DDM models discussed in Section 1 as they allow us to build a bridge to psychological factors and brain dynamics (see also \cite{Ratcliff2016,Zhu2023,Saraiva2023,Castagna2023}, a recent review \cite{Myers2022}, and references therein). In the following sections, we provide further details on how sensory observations and subsequent learning via brain networks can be linked to the model discussed here. 

	\subsection{Bayesian models for decision-making}
	
	An intrinsic relationship between probabilistic drift-diffusion and Bayesian models has been emphasized in neuroscience literature for quite some time now, with a strong advocacy for their applications in the modelling of decision-making and learning processes, including reinforced learning (see, e.g., \cite{Wiecki2013,Pedersen2017}).
	One of the critical class of Bayesian models is the Bayesian model for concrete sensory observations. To recognize a
	presented stimulus a Bayesian model compares predictions, based
	on a generative model, to the observed sensory input. Similar to brain networks such generative models include certain distribution of the data itself. Through
	Bayesian inference, this comparison leads to belief values indicating  how probable it is that the stimulus caused the sensory
	observations. Note that this is conceptually different from the
	DDM where the decision process accumulates random pieces
	of evidence and there is no explicit representation of raw sensory input \cite{Gold2007,Bogacz2006,Bitzer2014,Fard2017}. Therefore, a combination of these modelling approaches can be beneficial in practice (e.g., \cite{Boelts2022}). 
	
	A Bayesian model is more complex than the probabilistic DDM. There are 4 required components \cite{Bitzer2014}:
	\begin{itemize}
		\item[(i) ] \quad The generative input process (reflecting the physical environment) which generates noisy observations of stimulus features just as those used in the actual experiment.
		\item[(ii) ] \quad The internal generative models of decision makers which mirror the generative input process under the different, hypothesized decision alternatives. 
		\item[(iii) ] \quad The inference mechanism which translates observation from (i) into posterior beliefs over the correctness of the alternatives using the generative models (ii). 
		\item[(iv) ] \quad A decision policy that makes decisions based on the posterior beliefs from (iii). 
	\end{itemize}
Bayesian models can be extended to include fractional Brownian motions \cite{Thapa2022}, in which case an extension of model \eqref{DDM}-\eqref{stoping_time} would also be required. Other extended DDMs for decision-making and learning have also been proposed (e.g. \cite{Fengler2022}).  
	\subsubsection{Input process and observational sensory information for decision-making}
	
	In the brain, the sensory observations are reflected by the input process. In particular, sensory observations such as visuals, are reflected in an input translated into simple, noisy feature values used for decision-making. Assume that the observational sensory processes are drawn from a Gaussian distribution whose parameters we will infer from the behavioural data. In particular, we introduce the following input process with Gaussian distribution 
		\begin{align}
		X_t \sim \mathcal{N}(\mu_i, \Delta t \sigma^2), 
	\end{align}
where $\mu_i$ is the feature value which the brain would extract under perfect noise-free observation conditions. Here, $\Delta t \sigma^2$ is the variance representing the coherence of the dots (more significant variance equals smaller coherence) together with physiological noise in the brain.

While our better understanding of transforming sensory inputs into percepts represents one of the principal goals in neuroscience \cite{Manning2023}, the above framework assists us in formally integrating the knowledge with incoming sensory information. Knowledge creation is a complex process requiring an adequate mathematical framework, and the interested reader can consult \cite{Tadic2017} for further steps in that direction and a recent survey on related issues \cite{Tadic2021Self}. In what follows, this issue is addressed via the generative modelling approach under the assumption that the structure of internal representations in the brain replicates the design of the generative process by which the input process and observational sensory information influence it \cite{Ramstead2020}.
	\subsubsection{Generative models in Bayesian cognitive science}
	
	One of the key ingredients of applications of the free-energy principle to neuroscience and biological systems is active inference with a decisive role played by generative models \cite{Ramstead2020,Isomura2022}. The latter provides a guidance on how sensory observations are generated and how probability-density-based prior beliefs of a cognitive system (e.g., an individual or collective humans) about its environment and other information are controlled. Such generative models enter prominently the Bayesian framework in cognitive science where the underlying idea is that cognitive processes, including those playing the central role in decision-making, are underwritten by predictions based on inferential models \cite{Ramstead2020}. Assume that the decision maker aims to adapt its internal generative models to match those of the input process. We introduce the following generative model of an abstracted observation $X_t$ for an alternative $A_i$ as Gaussian densities
	\begin{align}
		p(X_t|A_i)= \mathcal{N}(\hat{\mu}_i, \Delta_t \hat{\sigma}^2),
	\end{align}
	where $\hat{\mu}_i$ is the mean, while $\Delta_t \hat{\sigma}^2$ represents the internal uncertainty of the decision maker's representation of its observations.
	
	This approach is frequently used in what is now termed as Bayesian neurophysiology \cite{Isomura2022} as it allows us  to empirically explain many important brain functions in terms of Bayesian inference. Its formal definition is given next.
	\subsubsection{Bayesian inference for decision-making processes}
	
	The active inference is one of the main components of the Bayesian models. In this Bayesian inference, there is a posterior belief $p(A_i| X_t)$ that alternative $A_i$ is true given observation $X_t$. In the perceptual decision-making process, where observations $x_t$ arrive sequentially over time, a key quantity is a posterior belief $p(A_i|X_{1:t})$ where $X_{1:t} = \{X_1, \ldots, X_t\}$ collects all observations up to time $t$ \cite{Bitzer2014}. Then, this posterior belief could be computed recursively over time using Bayesian inference as follows (see more detail in e.g. \cite{Bitzer2014}):
	
	\begin{align}
		p(A_i| X_1) &= \frac{p(X_1|A_i)p(A_i)}{\sum_{j=1}^{M} p(X_1|A_j)p(A_j)} \label{eq_bay_1}\\
		p(A_i| X_{1:t}) &= \frac{p(X_t|A_i) p(A_i| X_{1:t-1})}{\sum_{j=1}^{M}p(x_t|A_j)p(A_j|X_{1:t-1})}, \label{eq_bay_2}
	\end{align}
	where $M$ represents the number of considered alternatives. 
Here, the equations \eqref{eq_bay_1}-\eqref{eq_bay_2} imply that the posterior belief of alternative $A_i$ is calculated by weighting the likelihood
of observation $X_t$ under alternative $A_i$ with the previous posterior
belief and normalizing the result. At the initial time step the previous belief is the prior belief over alternatives $p(A_i)$ which can
computed biases over alternatives.

The development of the concept of active inference goes hand in hand with recent advances in neuroscience, allowing the characterization of brain functions based on mathematical formalisms and first principles. As a result, the application of this approach grows, including the areas critical for collective human interfaces and decision-making such as Neurorobotics and Artificial Intelligence (AI) \cite{Costa2022,Badcock2022,Kiverstein2022}.  
	\subsection{Decision policy for decision-making processes}
	
	In Bayesian models, decisions lie in the posterior belief $p(A_i| X_{1:t})$. Then, a decision is made for the alternative with the largest posterior belief when any of the posterior beliefs reaches a predetermined bound $\lambda$, which reads (see, e.g., in \cite{Bitzer2014}):
	\begin{align}
		\max_i p(A_i|X_{1:t}) \geq \lambda. 
	\end{align}
On the other hand, the decision variables with the posterior beliefs can also be describe by the following formula:
\begin{align}
	\max_i \log p(A_i|X_{1:t}) \geq \lambda',
\end{align}
where $\lambda'$ is the posterior belief bound, while $p(A_i|X_{1:t})$ is determined as in \eqref{eq_bay_2}. 

We have shown the general picture of Bayesian models for decision-making processes. In order to understand further the Bayesian inference in modelling decision-making processes, we are also interested in 
how the brain extracts information from the sensory input signal that leads to decision-making. Hence, we will discuss further this approach through an example in the next section, where we use the Bayesian inference method based on the generalized linear models to describe an escape route decision scenario.

 In what follows, we will provide a few representative examples which we will use to demonstrate the application of the theoretical framework discussed in the previous section. 
	\section{Examples}\label{sec:examples}
	\subsection{State-of-the-art in modelling risky decision-making}

People make risky decisions during fire evacuations such as moving through the smoke. Although a shortcut in a smoky area may help individuals evacuate quickly, it is still dangerous. A wrong decision in choosing the escape route during fire evacuations could lead to the risk of injuries or death \cite{Fudenberg2020,Xu2020,Iinuma2021}. These earlier studies investigated the effects of smoke levels, individual risk preference, and neighbor behaviour on individual risky decisions to take a smoky shortcut for evacuations. 

%
We note further that to respond to indoor fires, people often choose to evacuate from hazardous buildings to a safe place. Hence, evacuation route selection plays a critical role in determining the evacuation efficiency and whether evacuees can leave a hazardous area safely. In a perfect world, people would rationally avoid ongoing or imminent hazards when selecting a route to escape quickly. However, risk-taking behavior is widely observed during evacuations. In a building fire, people may be unaware of or underestimate the danger such as smoke and take a risky route for evacuations.
Taking a shortcut in a hazardous area such as a smoky corridor or stairs is a typical risk-taking behavior. In particular, in emergency situations, e.g. fire evacuations such as moving through smoke, humans could make decisions and move in a panic mood. Hence, psychological factors we discussed earlier in this chapter would be playing a crucial role. In such scenarios together with the high density of smoke, one may have uncertainty illusions such as right shortcut illusions. Let us define the so-called right-shortcut illusion when the participants feel that this shortcut is the right route to lead to the fastest evacuation. Motivated by \cite{Fu2021}, we consider the human decision-making process model in choosing smoky shortcuts during fire evacuations, such as moving through the smoke.

When individuals have to make decisions on whether to evacuate through smoke, they have uncertainty about the accessibility of the smoky route. Hence, individuals may treat neighbor behavior as useful information when making judgments on risky route choices. However, such social influence on individual risk-based decision-making still needs experimental investigation \cite{Hoffmann2013,Kwon2022}. In our consideration, we assume that humans have normal abilities in vision, color recognition, auditory sense, and movement. Often humans have the wrong percept \cite{Dokka2015}. In particular, they think their own route might be the best choice in escaping the emergency situations when the other neighbors might have other better choices of escape route; or vice versa. The illusion is usually resolved once you gain a vision of the surroundings that lets you disambiguate the routes.
We asked the following question: "How do noisy sensory estimates of vision lead to uncertainty percepts of choosing the right shortcut?"
In what follows, we provide representative numerical examples with DDM and Bayesian inference to address the proposed question. 
\subsection{Numerical results with DDM for a decision-making model}\label{DDMexample}

It is well known that there is a relationship between visualization perception and sensory cortex signals \cite{Dokka2015,Usher2001,Roy2021}. The major part of the brain's role is devoted to processing the sensory inputs that we receive from the world. Then, by generating spikes of activity, neurons in the sensory cortex respond to these stimuli receiving from the surrounding enviroments \cite{Singer2018}. In order to demonstrate the application of the theoretical framework discussed in the previous Section \ref{sec:DDMBayesian}, we provide numerical examples to investigate the visualization of surrounding environments and sensory processing that leads to perceptual decision-making in the human brain. 

%
%
Using the definition of the general DDM in \eqref{DDM}-\eqref{stoping_time}, we consider the following specific model of decision-making in the case of 2 alternative choices for choosing the smoky shortcut and the other route: 
\begin{align}\label{DDM_2}
	de(t) = -c e(t)dt + v(z)dB(t),
\end{align}
where $e$ is the accumulated evidence, $v$ is our sensory cortex input already containing the noise, $c$ is the leakage constant, while $B(t)$ represents the standard Brownian motion, as in Section 2. Note that a decision-making threshold represents the value of the decision-making variable at which the decision is made, such that an action is selected, marking the end of the accumulation of information \cite{Miller2013}. In our consideration, the decision-making threshold \textquotedblleft$\text{Thr}$\textquotedblright \ is equivalent to a boundary condition. In this model, for $i=1,2,3,\ldots$, the sensory cortex signal generator can be defined as follows (see, e.g., \cite{Singer2018}): 
\begin{align}\label{sensory_signal}
	dv(z) = 1000\gamma \delta(z(x_i) - z(x_{i-1}))dt + \sigma dB(t),
\end{align}
where $\gamma, \sigma$ are constants and $$z(x_i) = \frac{1}{1+e^{-2x_i}}.$$

In this subsection, we use a DDM to model decision-making in the case of 2 alternative choices for our choosing escape route scenario in \eqref{DDM_2}.
The numerical results reported in this subsection are obtained by using a discrete-time integration based on the Euler method implemented in Python. In particular, we use the open-source framework provided by Neuromatch academy computational neuroscience (https://compneuro.neuromatch.io/). 
\begin{figure}[h!]
	\centering
	\includegraphics[width=0.8\textwidth]{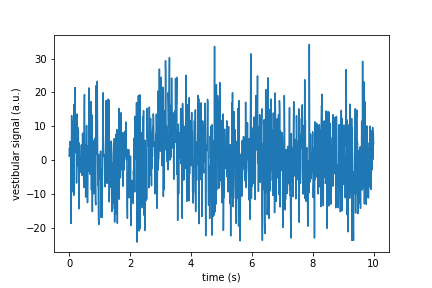}
	\caption{[Color online] Sensory cortex signal profile using the generator in \eqref{sensory_signal}.}
	\label{fig:0-1}
\end{figure} 
In what follows, we use the sensory cortex signal profile provided in Fig. \ref{fig:0-1} for all of our simulations. This sensory cortex signal has been generated by using the formula \eqref{sensory_signal}. The numerical results reported in this subsection are based on a simple but illustrative example compared to earlier results (e.g., \cite{Mormann2010,Bitzer2014,Masis2023}). Such results aim to investigate and get better insight into perceptual decision-making in the human brain. In this context, the DDM is a well-established framework that allows us to model decision-making for two alternative choices. At the same time, our complementary development of the Bayesian inference approach for these problems is more suitable for predicting such choices from spike counts of neurons. In more detail, we will discuss the Bayesian inference approach in Subsections \ref{Bayesianmodel}-\ref{sec:numBayesian}.

\begin{figure}[h!]
	\centering
	\includegraphics[width=0.8\textwidth]{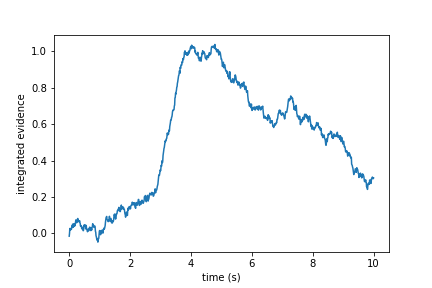}
	\caption{[Color online] Integrator profile (or drift-diffusion mechanism) by using the formula \eqref{DDM_2}.}
	\label{fig:0-2}
\end{figure}

\begin{figure}[h!]
	\centering
	\includegraphics[width=0.85\textwidth]{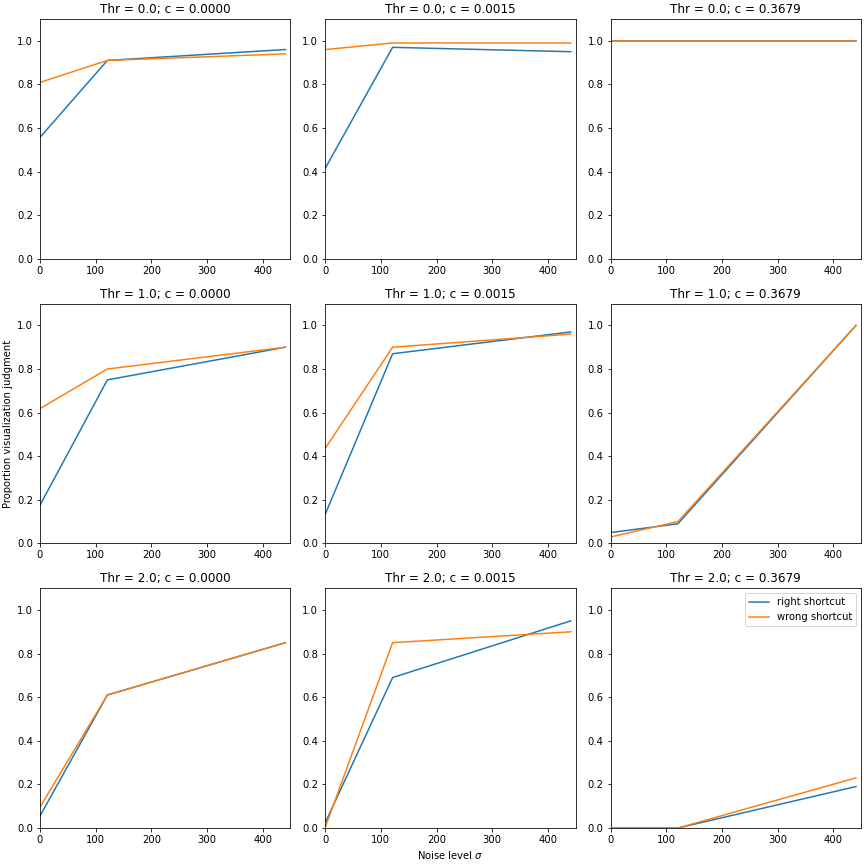}
	\caption{[Color online] Proportion escape route judgment as a function of noise. Blue and orange lines denote the right shortcut and wrong shortcut, respectively. First row: $\text{Thr} = 0$, $c= 0, 0.0015, 0.3679$. Second row: $\text{Thr} = 1$, $c= 0, 0.0015, 0.3679$. Third row: $\text{Thr} = 2$, $c= 0, 0.0015, 0.3679$. }
	\label{fig:0-3}
\end{figure}

\begin{figure}[h!]
	\centering
	\includegraphics[width=0.8\textwidth]{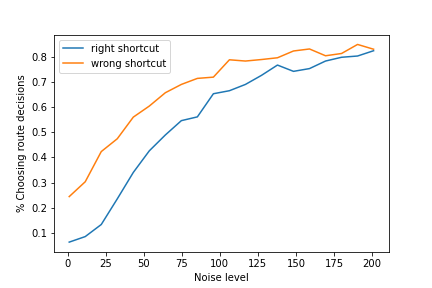}
	\caption{[Color online] Decisions on choosing smoky shortcut. Blue and orange lines denote the right shortcut and wrong shortcut, respectively. Parameters: $\text{Thr} = 1.5$, $c = 0.0004$. }
	\label{fig:0-4}
\end{figure}

The main numerical results of our analysis obtained with this DDM are shown in Figs. \ref{fig:0-2}-\ref{fig:0-4}, where we have plotted the integrator (or drift-diffusion mechanism), the proportion visualization judgment, and the choosing escape route decisions.
In particular, in Fig. \ref{fig:0-2}, we have plotted the drift-diffusion mechanism as the integrated evidence of our choosing escape route system. Here, we have the threshold equal to 1. 
Then, in Fig. \ref{fig:0-3}, we have plotted the proportion of escape judgment for our cognition of right escape route dynamics. These plots are used to evaluate and test our model \eqref{DDM_2} - \eqref{sensory_signal} performance. In particular, we test our hypothesis 1-2 proposed in the previous section for different parameter combinations. Then, we evaluate how our model behaves as a function of the 3 parameters the threshold $\text{Thr}$, leakage constant $c$ and noise level $\sigma$. We see that the presence of noise affects the proportion of escape route judgment. Moreover, an increase in the leakage  constant, together with an increase of threshold values, leads to the proportion escape route judgment of both right shortcut and wrong shortcut decrease. However, the results presented in Fig. \ref{fig:0-3} do not reflect exactly the properties of the right shortcut and wrong shortcut judgments. We will have a look at the following decisions on the right shortcut. 

In Fig. \ref{fig:0-4}, we have plotted the escape route decisions. Our numerical results show that our hypothesis of linear increase of visualization strength with noise only holds true in a limited range of noise. The percentage of right escape route decisions is higher than the wrong escape route decision. The curves presented in Fig. \ref{fig:0-4} are monotonic but saturating. 

Our numerical results in this subsection show that the noise pushes the integrated signal over the threshold. Additionally, we observe that the less leaky the integration and the lower the threshold, the more motion decisions we get. 

We have shown the DDM for the escape route visualization model. As we have mentioned in the Introduction and the model description sections, the DDM is also in connection with the Bayesian models. In order to analyze further how the brain extracts information from noisy input as typically presented in perceptual decision-making tasks. In what follows, we will consider another representative example, the decision-making process for the escape route decision phenomenon by using Bayesian inference approach.

\subsection{Bayesian inference modelling spiking neurons for decision-making processes }\label{Bayesianmodel}

As we have mentioned in the previous sections, collective decision-making can be described as the brain with a collection of neurons that, through numerous interactions, lead to rational decisions. The most commonly used tool to describe the stimulus
selectivity of sensory neurons is the generalized linear models (GLMs) \cite{Gerwinn2007,Macke2011}. Let us now recall the following class of GLMs, namely, the logistic regression model for predicting decision-making from spike counts. First, we introduce the fundamental input/output equation of logistic regression \cite{Kleinbaum2002}, which reads
\begin{align}\label{logistic}
	\hat{y} \equiv P(y=1|x,\theta) = \sigma(\theta^T x) = \sigma(z(\theta_i,x_i)), 
\end{align}
where $\hat{y}$ denotes the the output of logistic regression. Here, $\hat{y}$ can be considered as  the probability that $y = 1$ given inputs $x$
and parameters $\theta$. Additionally, $\sigma$ represents a "squashing" function called the sigmoid function or logistic function. The output of such logistic function is defined as follows: 
\begin{align}
	\sigma(z(\theta_i,x_i)) = \frac{1}{1 + e^{-z(\theta_i,x_i)}},
\end{align}
where $z(\theta_i,x_i) = \alpha + \theta_1x_1 + \theta_2x_2 + \ldots +\theta_n x_n = \sum_{i=1}^{n}(\alpha + \theta_i x_i)$, for $i \in \mathbb{N}$.
Motivated by \cite{Gerwinn2007,Weber2017}, we are interested in a Bayesian treatment of the models for predicting stimulus from spike counts for our decision-making processes. Our methodology can be generalized to other classes of models that go beyond the GLM class, described above (e.g., \cite{Theis2013,Fortuna2023}).

In this subsection, we investigate the sensory evidence accumulation activity during human decision-making. 
We have built the risky decision model presented in subsection \ref{DDMexample}. That model predicts that accumulated sensory evidence from sensory cortex signals determines whether the human should choose the smoky shortcut. Here, using the descriptions of Bayesian inference, we will build the sensory neuron data and would like to see if that prediction holds true. 

The data contains $N=40$ neurons and $M=400$ trials for each of the three visualizing conditions: no smoky shortcut, slightly smoky shortcut and high-density smoky shortcut.

In order to address our question, we need to design an appropriate computational data analysis pipeline. Moreover, we need to somehow extract the escape route judgements from the spike counts of our neurons. Based on that, our algorithm needs to decide: was there a right shortcut or not? This is a classical two-choice classification problem. We must transform the raw spike data into the right input for the algorithm (the process known as the spike pre-processing, e.g., \cite{Anumula2018}). 



 Noise in the signal drives whether or not people perceive visualization of the smoke level. The brain may use the strongest signal at a peak level of noise to decide on choosing the shortcut, but we actually might think it is better to accumulate evidence over some period of time. We want to test this. The noise idea also means that when the signal-to-noise ratio is higher, the brain does better, which would be in the high density of smoke condition. We want to test this too.

Using the description of logistic regression \cite{Kleinbaum2002}, as an example, we introduce the following hypotheses focussing on specific details of our overall research question:

Hypothesis 1: Accumulated sensory cortex spike rates explain visualization of smoke judgements better than average spike rates around the peak of the smoke level, and

Hypothesis 2: Classification performance should be better for high-density smoke shortcuts and low-density smoke shortcuts.

Mathematically, we can write our hypotheses as follows (using our above ingredients):
\begin{itemize}
	\item Hypothesis 1: $E(c_\text{accumulate}) > E(c_\text{average spike})$;
	\item Hypothesis 2: $E(c_\text{high smoke density}) > E(c_\text{low smoke density})$,
\end{itemize}
where $E$ denotes taking the expected value (in this case the mean) of its argument: classification outcome in a given trial type.

In what follows, we use the logistic regression as a case in point to predict stimulus from spike counts for the proposed escape route decision problem. 

\subsection{Numerical results with the Bayesian approach for a decision-making model}\label{sec:numBayesian}

In this subsection, we use a Bayesian inference approach to model decision-making in the case of 2 hypotheses provided in the previous subsection \ref{Bayesianmodel} for our choosing the right escape route scenario.
The numerical results reported in this subsection are obtained by using a logistic regression model, as an example, implemented in Python. In particular, we use the open-source framework provided by Neuromatch academy computational neuroscience (https://compneuro.neuromatch.io/). We are also using the method of logistic regression to predict stimulus from spike counts. As mentioned earlier, other models that go beyond the GLM class can also be used in this context.

\begin{figure}[h!]
	\centering
	\includegraphics[width=0.65\textwidth]{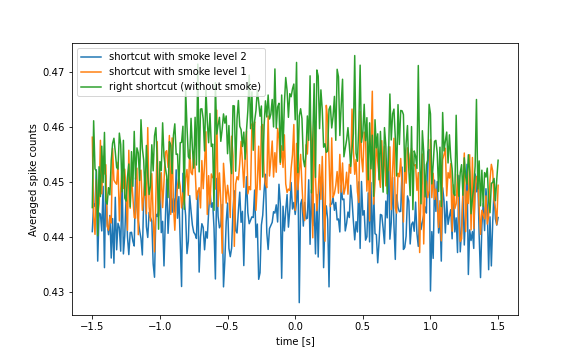}
	\caption{[Color online] The time evolution of the averaged spike counts for the escape route model provided in subsection \ref{Bayesianmodel}. Blue, orange and green lines represent the averaged spike counts of shortcut with smoke level 2, shortcut with smoke level 1 and right shortcut, respectively.}
	\label{fig:2}
\end{figure}

\begin{figure}[h!]
	\centering
	\includegraphics[width=0.65\textwidth]{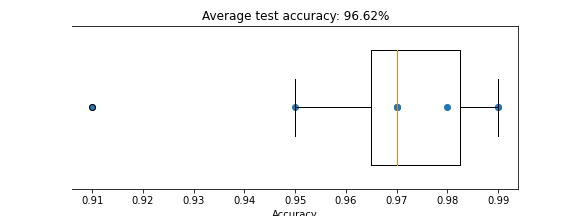}
	\caption{[Color online] The average test accuracy profile obtained by using classifier accuracy of the logistic regression (see, e.g., \cite{Cerulli2022,Nelli2023}). Blue dots denote the accuracy values, the orange line crossing a blue dot represents the average accuracy at that point.}
	\label{fig:5}
\end{figure}

The main numerical results of our analysis in this Bayesian approach are shown in Figs. \ref{fig:2}-\ref{fig:5}, where we have plotted the average spike counts. The average test accuracy profile and the comparison of the accuracy between the right shortcut and right shortcut judgements. 

In Fig. \ref{fig:2}, we have plotted the averaged spike counts. Blue represents the shortcut with a high density of smoke condition and produces flat average spike counts across the 3s time interval. The orange and green lines show a bell-shaped curve corresponding to the smoke level profile. There are fluctuations in the averaged spike counts. It is clear that there is noise in our consideration. In order to see the effects of noise on the data accuracy, we look at the following results in Fig. \ref{fig:5} on the average test accuracy profile (see, e.g., \cite{Roy2021,Wang2021neural}). 
%

In Fig. \ref{fig:5}, we have plotted the average test accuracy profile obtained by using classifier accuracy of the logistic regression. Prediction accuracy ranges from 91\% to 99\%, with the average at 95\%, and the orange line is the median at 97\%. We observe that our prediction has a high accuracy even though the given data includes noise factors. 
It could be better to split the average accuracy according to the conditions where we could visualize the escape route but of different magnitudes. Then, it should work better to classify higher smoke density from no smoke as compared to classifying the lower smoke density. 

The spike activity also works better if we ignore some of the noise at the beginning and end of each trial by focusing on the spikes around the maximum smoke density, using our window option. Additionally, we see that the average spike count plot above seems to best discriminate between the three levels of the smoke density conditions around the peak at time zero.
\begin{figure}[h!]
	\centering
	\includegraphics[width=0.65\textwidth]{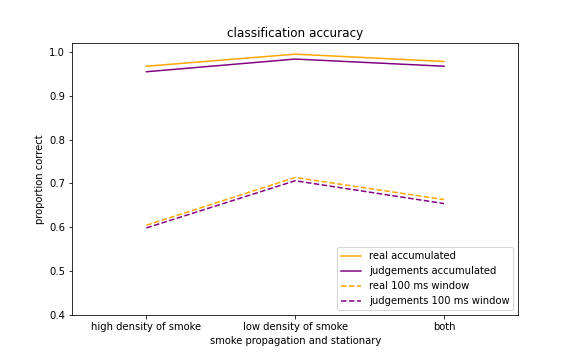}
	\caption{[Color online] The average classification accuracy profile for total accumulated spikes and for a window around the peak of the spike counts (see, e.g., \cite{Cerulli2022,Nelli2023}). Orange and purple lines denote the real accumulated spikes and the juidgements accumulated spikes. Orange and purple dash lines represent the real 100 ms window and judgements 100 ms window.}
	\label{fig:5-2}
\end{figure} 

Looking at Fig. \ref{fig:5-2}, using the logistic regression, it is clear that the results for small window data of 100 ms and the full data set are totally different. In particular, the accuracy between the true escape route and wrong escape route judgements in the case of a small window of data is smaller than in the case of a full data set. This is due to the fact that the brain also integrates signals across longer time frames for perception. On the other hand, in the predictions based on accumulated spike counts, the high density of smoke are harder to separate from no smoke than the low density of smoke. This is clearer when predicting real choice than when predicting escape route judgements. Moreover, it is also clear that the real accumulated spike counts approximate the judgements accumulated spike counts for small window data and for the case of a full data set. We observe that the logistic regression works quite well in the case of solo small window data and the case of a solo full data set. 


Notice that right escape route judgments display higher decoding accuracy than the wrong choice. If right escape route judgements and our logistic regression use input from the same noisy sensors, it can be expected that they would both give similar output. This aligns with the notion that escape route judgements can be wrong because the underlying sensory signals are noisy. Of course, this only works if we record activity from neuronal populations contributing to escape route judgements. On the other hand, we would also see this result if the sensory signal was not noisy and we recorded from one of several populations that contribute to escape route judgements in a noisy way.

We have shown representative examples of DDM and Bayesian inference models for human decision-making. The DDM approach has described the decision-making of individuals, while the Bayesian inference, based on logistic regression or other models, could describe  collective decision-making via the brains with a collection of neuron activities. Both approaches lead to human decisions in different views. However, we know that humans not only interact individually but also interact with human groups. The decision-making problems in this situation bring a lot of challenging questions to scientists due to the complexity of human behaviour. Taking this inspiration, in what follows, we will provide a review of the recent developments in  collective human decision-making and its application in brain network models. 

Our numerical simulations can be compared with the numerical results reported in \cite{Fu2021}. Using an immersive virtual reality (VR)-based controlled experiment, the authors in \cite{Fu2021} have studied the effect of smoke level, individual risk preference, and neighbour behaviour on individual risky decisions to take a smoky shortcut for evacuations. The study in \cite{Fu2021} aimed to conduct a controlled experiment to verify the influence of the smoke, individual risk preference, and neighbour behaviour on individual risky decisions, i.e., whether to evacuate through the smoke. The experiment manipulated the density of the virtual smoke in the immersive virtual environments to investigate the effect of smoke level on participants’ route selection (see, e.g., in \cite{Fu2021} and related references therein). Their numerical results showed that a higher smoke density lowered the use rate of a smoky shortcut during evacuations when participants needed to choose between the risky shortcut and another available route without the smoke. In particular, 89.05 \% of participants evacuated through the shortcut, but the percentage reduced to 55.24\% in the slight smoke scenarios, with 25.24\% in the cases of dense smoke. In \cite{Fu2021}, the authors considered the influence of smoke, individual risk preference, and neighbour behaviour on individual risky decisions. However, our model investigates the simpler case of only smoke effects from the computational neuroscience perspective. We also found that the density level of the smoke is a critical factor in determining whether people will take a risk. When the smoke density increases, humans tend to be willing not to choose a smoky route. Next, we will devote our efforts to deeper exploring intrinsic links between collective decision-making and the complex operation of brain networks.


\section{Collective decision-making and brain networks}\label{sec:4collective}

One of the most important topics in human decision-making studies is collective decision-making. This topic has attracted the interest of a large number of scientists from different fields, including mathematicians, engineers, psychologists and neuroscientists. Collective decision-making has been explained as a fundamental cognitive process required for group coordination \cite{Reina2021}. In order words, this process can be considered as the subfield of collective
behaviour concerned with how groups reach decisions \cite{Bose2017}.  The group decision-making processes can account for also the unavoidable variation in individual decision accuracy. Decision theory has been applied successfully to human groups by showing how
to optimally weight individuals’ contributions to the group
decisions according to their accuracy. The authors in \cite{Teodorescu2016} have provided empirical evidence for human sensitivity to task-irrelevant absolute values indicating a hardwired mechanism that precedes executive control. On the other hand, the collective decision-making processes and social learning processes, including opinion dynamics, are closely connected in social science. In particular, many researchers investigate collective decisions in humans, deepening our understanding of the dynamics of economies and social policies \cite{Galam2008,Molavi2018,Groeber2014,Mavrodiev2021,Rauhut2011}. A multi-agent system is also a collective of autonomous agents interacting in a shared environment  \cite{Barfuss2022}. The multi-agent systems play an important role in a variety of application domains, such as traffic, human decision-making, control, and complex systems \cite{Adler2002,Melnik2009coupling,Bulling2014,Geng2020}. They appear to be indispensable tools for studying bio-social interactions and play an important role during the recent pandemic (e.g. \cite{Tadic2020,Tadic2021}). An agent-based model to explain the emergence of collective opinions not based on feedback between different opinions, but based on emotional interactions between agents has been proposed in \cite{Schweitzer2020}.

Collective decision-making is described not only as individuals in a group either reaching a consensus on
one of several available options or distributing their workforce over different tasks but also as the brains with a collection of neurons that, through numerous interactions, lead to rational decisions \cite{Bogacz2006,Bose2018,Park2019,Reverberi2022}. In \cite{Gold2007}, the authors have evaluated recent progress in understanding how these basic
elements of decision formation, including deliberation and commitment, are implemented in the brain. In particular, the decisions are characterized by many sensory-motor
tasks that can be thought of as a form
of statistical inference. Additionally, we know that many aspects of human perception are best explained
by adopting a modelling approach in which experimental subjects are assumed to
possess a full generative probabilistic model of the task they are faced with and
that they use this model to make inferences about their environment and act
optimally given the information available to them \cite{Vijayakumar2011}. Hence, these decision-making systems normally include noise. A number of researchers in \cite{Gold2007,Teodorescu2016} have addressed the question, "What is the (unknown)
state of the world, given the noisy data provided
by the sensory systems?".  The elements of such a decision-making process are described in terms of probability
theory, e.g. Bayesian methods \cite{Doya2007,Gold2007}. Recently, the Bayesian approach to perception and action has been used in modelling human decision-making. This approach has attracted the interest of many researchers from different fields and has successfully accounted for many experimental findings \cite{Acerbi2014,Ma2023}. Unlike the individual decision-making proposed in the previous sections, such Bayesian inference can be also used to model collective decision-making. One of the challenging questions the brain research accounted for regarding human social decision-making is when decisions are made in a social context, the degree
of uncertainty about the possible outcomes of our
choices depends upon the decision of others. A model-based account of the neurocomputational mechanisms
guiding human strategic decisions during collective decisions has been discussed in \cite{Park2019}. An influential review of brain theories in the biological (for example, neural Darwinism) and physical (for example, information theory and optimal control theory) sciences from the free-energy perspective has been given in \cite{Friston2010}. Using the free-energy principle and active inference approach, two free-energy functionals for active inference in the framework of Markov decision processes have been compared in \cite{Parr2019}. In developing an optimal Bayesian framework based on partially observable Markov decision processes, the authors in \cite{Khalvati2019} have shown that humans simulate the “mind of the group” by modelling an average group member’s mind when making their current choices, in the group decision-making. A brain network supporting social influences
in human decision-making has been discussed in \cite{Zhang2020}. Such social influences can often lead to interactive decision-making under partially available information \cite{Kumar2022}. One of the important classes of collective decision-making is the self‑driven collective dynamics with graph networks \cite{Tarcai2011,Wang2022}. This collective dynamics plays an important role in self-organization for decision-making processes \cite{Valentini2014}. 
A central concept connecting the microscopic and macroscopic levels of neurons is criticality in brain studies \cite{Hesse2014}. Moreover, the main elements of the criticality hypothesis are the evolutionary arguments and a plausible general mechanism that can explain the self-organization to the critical state \cite{De2006,Hesse2014}. A review of the experimental findings of isolated, layered cortex preparations to self-organize toward
four dynamical motifs presently identified in the intact cortex in vivo: up-states, oscillations,
neuronal avalanches and coherence potentials have been provided in \cite{Plenz2021}. 



\begin{figure}[h!]
	\centering
	\includegraphics[width=0.85\textwidth]{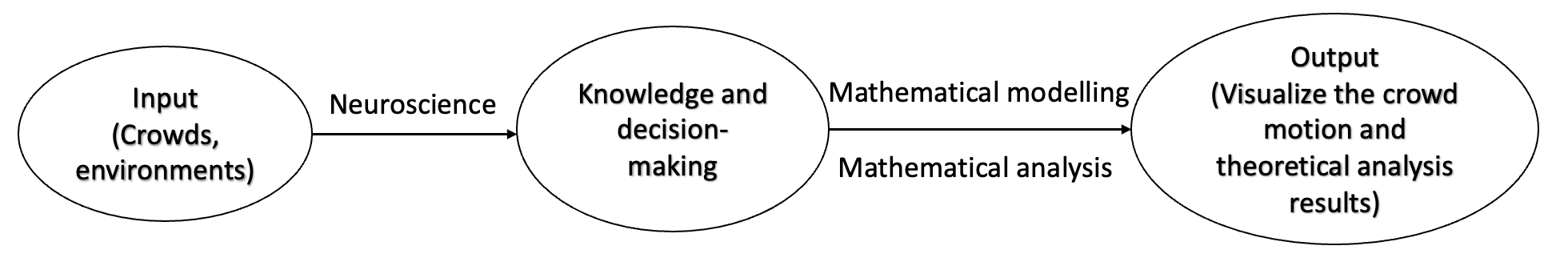}
	\caption{
		{\small Schematical illustration of the human decision-making dynamic modelling. Humans receive information from the surrounding environments. The evidence comes
			from sensory input, which in this work is formulated to
			reflect the visual information the humans receive from
			emergencies (e.g., fire, earthquakes, etc). Then, a decision is made when the value of the diffusion process reaches a decision threshold. Finally, the output will be the visualization of the crowd motion and theoretical analysis results. This illustration demonstrates bridging the gap between neuroscience and mathematical modelling and analysis (see references in the text). }}
	\label{fig:0}
\end{figure}

There are many ways to highlight different features of human decision-making dynamic modelling and a rich set of associated mathematical problems — one of them we present in Fig. \ref{fig:0}. In this figure, we highlight the importance of the triad: neuroscientific foundations, mathematical modelling, and analysis.
In particular, starting with neuroscientific considerations, we use mathematical modelling to build the models for this complex process. Then, we use mathematical analysis to theoretically prove the well-posedness and show the properties of the associated models. In bridging the gap between the different components of the above triad and addressing related problems, some progress has been achieved (e.g.,  \cite{Zhang2021,Boehm2021,Boehm2022,Thieu2022coupled} and references therein), with many open issues remaining. Note also that collective decision-making is not limited to the human behavioural system. It is ubiquitous across the living and artificial collectives \cite{Reina2021}. 

%

Collective dynamics also include collective emotions. Many models have been proposed to capture collective emotions \cite{Sander2020,Brown2021,Goldenberg2020}. Individual and group-based emotions
are individual-level or micro-level phenomena. In contrast, collective
emotions are defined as macro-level phenomena that
emerge from emotional dynamics among individuals
responding to the same situation \cite{Goldenberg2020}. In \cite{Goldenberg2020}, the authors have discussed collective emotions in the larger context of
collective-level psychological phenomena, defined collective
emotions and discussed their key components, and
then showed how collective emotions emerge from individual-level emotional interactions. On the other hand, there is another direction in capturing the collective emotion dynamics. For instance, the collective emotional dynamics during the COVID-19 outbreak have been considered in \cite{Metzler2022}. Recent results on learning dynamics with graph networks are provided in \cite{Wang2022}, where the authors have shown another interesting direction in capturing the learning of self‑driven collective dynamics with graph networks. The proposed approach could potentially be useful in modelling collective decision-making by using learning dynamics with graph networks. 

Understanding the character of the dynamics of sensory decision-making behaviour offers many challenging questions to scientists due to the fact that decisions may depend on a large number of task covariates, including the sensory stimuli, an agent’s choice bias, past stimuli, past choices, past rewards, etc. In what follows, in order to understand better collective human decision-making, we provide numerical examples of collective dynamics in the approach based on brain networks considered as collections of neurons as well as in the group dynamics. 

	\section{Examples of collective dynamics in the approach based on brain networks considered as collections of neurons}\label{sec:examplescollective}
	
	A large part of brain regions is critically involved in solving the problem of action selection or decision making, e.g. the cortex and the basal ganglia \cite{Bogacz2007TheBG,Zhang2020}. Furthermore, neuromodulation systems also participate in a variety of cognitive processes,
such as motivation, mood, and learning \cite{Bang2020,Marzecova2021,Grossman2022}.
Take dopamine as an example of neuromodulation. Dopamine's role is one of the most important factors in reward processing and motivation in decision-making. Many researchers from different fields have investigated
strong evidence of the role of dopamine in learning the value of actions, stimuli and states of the environment \cite{Montague2004,Bang2020,Iglesias2021}. In the basal ganglia (BG), in particular to the striatum, but also to the frontal cortex, the substantia nigra is also a crucial source of dopamine \cite{Sojitra2018,Oettl2020}. A series of mechanisms that reinforce and favor stimuli and actions are implemented by the dopaminergic system \cite{Belkaid2020}.

	 Understanding what drives changes in decision-making behavior is in the domain of reinforcement learning (RL) \cite{Botvinick2012,Dabney2020}. RL is a framework for defining and solving a problem where an agent learns to take actions that maximize reward \cite{Rmus2021,Eckstein2021}. The main feature of RL is that it explicitly considers the whole problem of a goal-directed agent interacting with an uncertain environment. 
The main elements of RL include an agent, biological or artificial, that observes the current state of the world and selects an action based on that state. In particular, while taking action, the agent receives a reward and then uses this information to improve its future actions. 
	
\begin{figure}
	 	\centering
	 	\begin{tabular}{lll}
	 		\includegraphics[width = 0.32\textwidth]{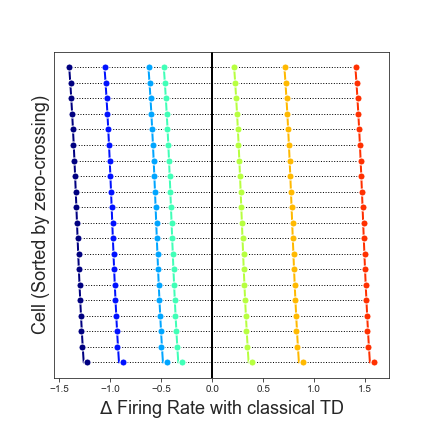} &
	 		\includegraphics[width = 0.32\textwidth]{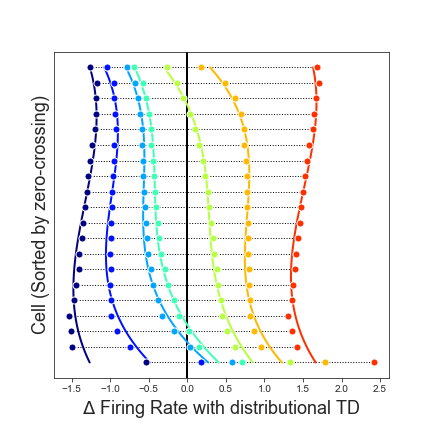} &
	 		\includegraphics[width = 0.32\textwidth]{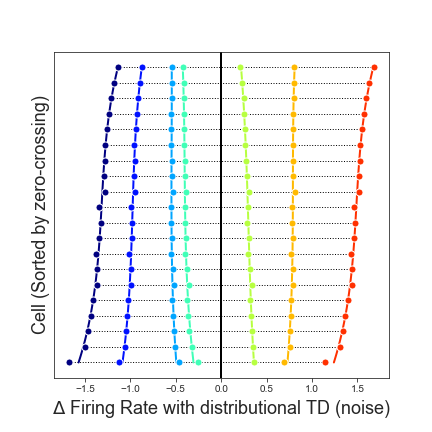}
	 	\end{tabular}
	 	\caption{\small [Color online] Different dopamine neurons profile. In this simulation, we ran 50 trials for 200 dopamine cells of 25,000
	 		updates with random learning rates in $[0.001, 0.02]$. On each trial, the human experiences one of seven possible reward magnitudes are selected randomly, e.g. 0.1, 0.5, 1.5, 2., 5.5, 9.5, 18 $(\mu l)$. First panel: RPEs produced by classical TD simulations. Second panel: RPEs produced by distributional TD simulations. Last panel: RPEs produced by distributional TD simulations with noise. Each horizontal bar is one simulated neuron. Each dot colour corresponds to a particular reward magnitude. The $x$ axis is the cell’s response (change in firing rate) when the reward is delivered. Cells are sorted by reversal point. In the last panel, the firing rate profile has different behaviour compared to the classical TD and the distributional TD. The firing rate increase dramatically, and the cells carry approximately the same RPE signal in the case presented in the last panel.}
	 	\label{fig:6}
	 \end{figure}

	 
	The action sequences of human decision-making usually involve many cognitive processes such as beliefs, desires, intentions, and theory of mind, i.e., what others are thinking. Due to the complexity of human behaviours, artifical intelligence (AI) as a powerful tool that predicts human behaviours to be treated agnostically to the underlying psychological mechanisms \cite{Lin2022}. The developments of AI algorithms achieve the higher-level brain-inspired functionality studies \cite{Hassabis2017,Chen2022}. One of the most important classes of RL in the field of AI is the reward prediction error. Based on the recent work in \cite{Dabney2020}, we recall a model of dopamine-based reinforcement learning inspired by recent artificial intelligence research on distributional reinforcement learning. Temporal difference (TD) is a class of model-free RL methods which learn by bootstrapping from the current estimate of the value function \cite{Sutton2018}. Unlike classical TD learning, we introduce briefly the distributional TD method. For the observed $x$, let $f: \mathbb{R}$ to $\mathbb{R}$ be a response function and a set of value predictions $V_i(x)$. We also have the set of values updated with learning rates $\alpha_i^{+}, \alpha_i^{-} \in \mathbb{R}^{+}$ to obtain the state $x'$ from the given state $x$. This process results in reward signal $r$ and the time discount $\gamma \in [0,1)$. The distributional TD errors are computed as follows:
	\begin{align}
		\delta_i = r + \gamma V_j(x') - V_i(x),
		\end{align}  
	where $V_j(x')$ represents a sample from the distribution $V(x')$. Then, the distributional model TD updates the baselines with the following fomula:
	\begin{align}
		V_i(x) \longleftarrow V_i(x) + \alpha_i^{+} f(\delta_i) \text{ for } \delta_i > 0,\\
		V_i(x) \longleftarrow V_i(x) + \alpha_i^{-} f(\delta_i) \text{ for } \delta_i \leq 0.
	\end{align}
	  The main numerical results have been obtained here using the open-source framework provided by \cite{Dabney2020} to simulate different properties of dopamine neurons with reward prediction error (RPE) theory. In particular, we apply the tabular simulations of the classical TD and distributional TD using a population of learning rates selected uniformly at random for each cell to obtain Fig. \ref{fig:6}. When we consider the difference between classical TD and distributional TD, we use a separately varying learning rate for negative prediction errors. The method use a linear response function. Moreover, the simulations for classical TD focus on immediate rewards, while the simulations for distributional TD learn distributions over multi-step returns. 
	  
 In Fig. \ref{fig:6}, we plot the different dopamine neurons that consistently reverse from positive to negative responses at different reward magnitudes. In the first panel, we plot the RPEs produced by classical TD simulations; we see that all cells carry nearly the same RPE signal. The presence of Gaussian noise causes slight differences between cells. Note that in contrast to classical TD learning, distributional
TD consists of a diverse set of RPE channels, each of which carries a different value prediction, with varying degrees of optimism across channels \cite{Dabney2020}. In the second panel, in the distributional TD case, cells have reliably different degrees of optimism. Unlike the classical TD, there are fluctuations in the system; we observe that some  cells use different RPE signals. However, we add Gaussian white noise to the distributional system, and the cells have their behaviours quite similar to the case of classical TD. Here, we observe that all cells carry nearly the same RPE signal. 
Using the assumption with the brain represents possible future rewards not as a single mean but as a probability distribution. All dopamine neurons should transmit essentially the same RPE signal (see more details in, e.g. in \cite{Dabney2020}). We have provided an example based on the results reported in \cite{Dabney2020}, where we have added the Gaussian white noise in the distributional TD case. We also observed that all cells carry nearly the same RPE signal. The presence of noise in the system could bring also benefits \cite{Faisal2008} since all dopamine neurons should transmit essentially the same RPE signal. The results provided in \cite{Dabney2020} have contributed to the development of the theory of dopamine. The study of dopamine provides explanations on a unifying framework for understanding the representation of reward and value in the brain. This direction of research would potentially be useful in the study of collective decision-making via the brains with a collection of interacting neuron systems provided in Sections \ref{sec:DDMBayesian}-\ref{sec:examples}. Additionally, a better understanding of uncertainty factors in dopamine-based RL could contribute to further developments in the fields of learning dynamics in human decision-making studies, brain disorders and other applications. We know that human decision-making is affected directly by learning and social observational learning. The models of interaction of direct learning and social learning at behavioral, computational, and neural levels have attracted the interest of researchers from different fields. The authors in \cite{Zhang2020} have provided a brain network model for supporting social influences in human decision-making. In particular, they have used a multiplayer reward learning paradigm experiment in real time. The numerical experiment shows that individuals succumbed to the group when confronted with dissenting information, but observing confirming information increased their confidence. Furthermore, the results could potentially contribute to the development of the study of learning dynamics in neural networks or social network problems. After appreciating the learning and social influence in decision-making, we turn to the social decision-making dynamics in groups. 

\begin{figure}
	\centering
	\includegraphics[width = 0.8\textwidth]{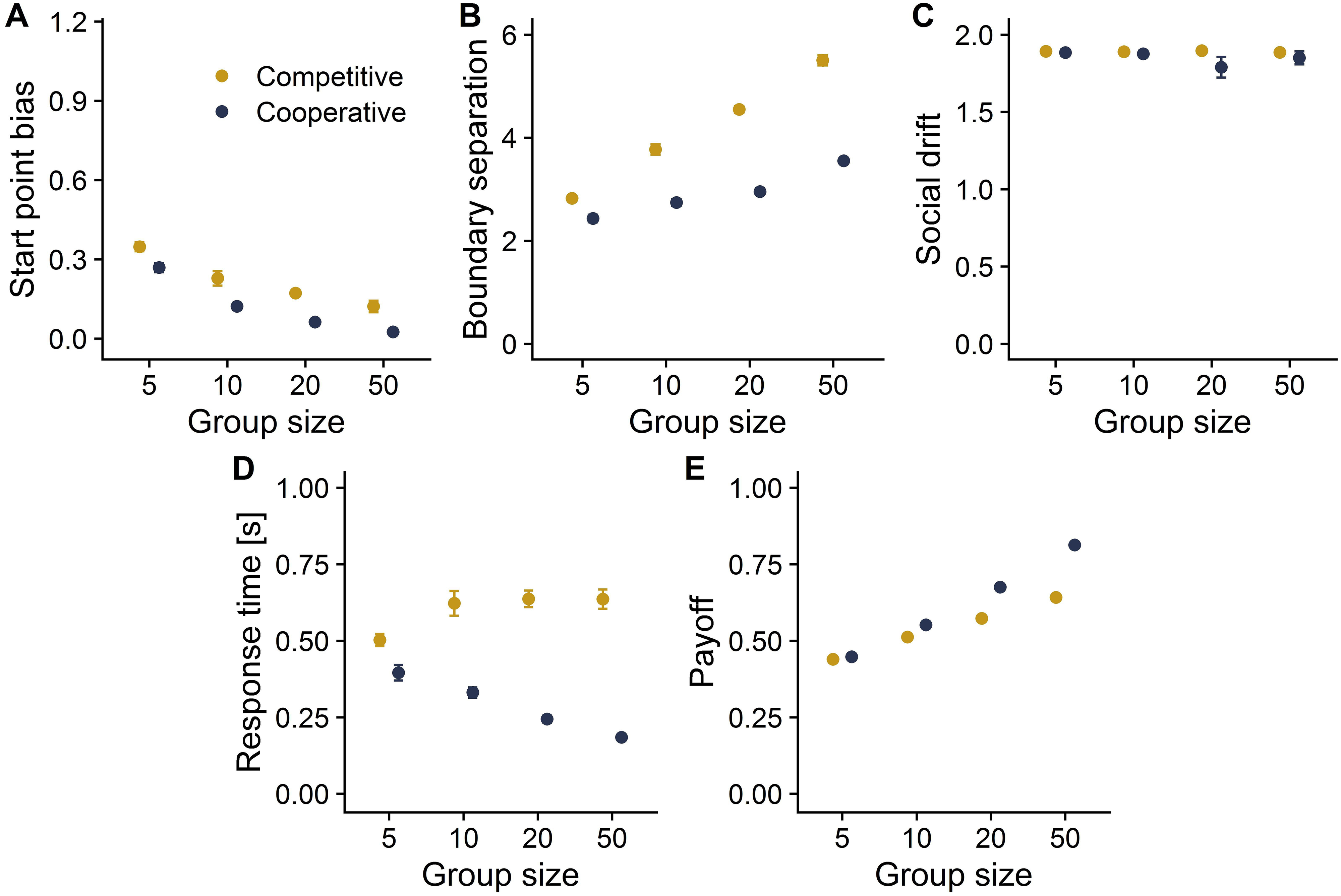}
	\caption{\small Reprinted by permission from \cite{Tump2022}. Copyright from the Creative Commons Attribution License (2022). Evolutionary outcomes of cooperative and competitive groups at an error cost ratio of 4, by \href{https://journals.plos.org/ploscompbiol/article?id=10.1371/journal.pcbi.1010442}{Alan N. Tump}, licensed under \href{https://creativecommons.org/licenses/by/4.0/}{CC by 4.0}. Across all group sizes, competitive groups evolved
(A) a larger start point bias and (B) larger boundary separation, indicating a conflict between individual- and group-level interests. (C) Both cooperative
and competitive groups evolved the maximum strength of the social drift. (D) Cooperative groups made, on average, faster choices than competitive
groups, and this difference increased with group size. (E) At large, but not small, group sizes, cooperative groups outperformed competitive groups.
Dots and error bars represent the mean and standard deviation of the endpoints of the evolutionary simulations, respectively.}
	\label{fig:9}
\end{figure}

Along with studying the decision-making of each individual as highlighted in Sections \ref{sec:DDMBayesian}-\ref{sec:examples}, the decision-making in groups of individuals is also important \cite{Gallup2012,Bragge2017}. Let us recall results reported in \cite{Tump2022}, where the authors have considered the decision process and information
flow with a DDM extended to the social domain. This DDM can be considered an extended version of the DDM provided in Sections \ref{sec:DDMBayesian}-\ref{sec:examples}. In particular, the implementation of the social DDM proposed in \cite{Tump2022} can be described as follows: Each individual first accumulates their personal
information about the state of the world. Then during the social phase, the individuals can account for additional social information, e.g. they can incorporate the choices of others. After an individual has sufficient evidence (i.e., the decision boundary is exceeded), the decision is
made. Applying the adaptive behavioural parameters and using evolutionary algorithms (see, e.g., \cite{Hamblin2013}), we can examine how individuals should strategically adjust decision–making traits to the
environment. A typical procedure for the analysis is as follows. First, we consider groups of individuals whose interests were completely aligned, with individuals
equally sharing a group payoff ("cooperative groups"). Then, we investigate whether
the collective interest was at odds with individuals’ self-interest. We 
examine how to introduce individual-level competition (i.e., a payoff solely based on own
performance) shaped evolved behaviours and corresponding payoffs across group sizes and
error cost asymmetries. To understand better the collective dynamics of such DDM, we look at the following recently reported numerical results presented in Fig. \ref{fig:9} (see \cite{Tump2022} for further details).

 Based on Fig. \ref{fig:9}, we can analyse the evolutionary outcomes of cooperative and competitive groups at an error cost ratio of 4, following the results first presented in \cite{Tump2022}. In panel (A), we see that competitive
groups developed a stronger start point bias towards the signal boundary than 
cooperative groups. In panel (B), the competitive groups evolved higher boundary separations than the cooperative
groups. On the other hand, we can also observe that both cooperative and competitive groups evolved to the maximum level of social drift
strength in panel (C). This observation confirms the strong benefits of using social information or even copying the first
responder, independent of group size or cooperative setting in the decision-making system. Additionally, in panels (D)-(E), we see that individuals in competitive groups made slower choices and achieved a lower payoff than individuals in cooperative groups. On the other hand, at larger group sizes, the cooperative and competitive groups achieved a
higher payoff. Moreover, cooperative groups have benefited much more from larger groups in panel (E).  This is due to the fact that the larger start point bias and boundary separation that
evolved in competitive groups partly undermined the benefits of collective decision-making (see more details, e.g., in \cite{Tump2022}).

The new and forthcoming results in this field, such as those reported in \cite{Tump2022}, could further contribute to understanding a wide range of social dynamics applications from crowd panics, medical emergencies and epidemics to critical law enforcement situations and smart city system designs.

	 As we have to deal with many challenges of uncertainty in collective human decision-making problems, some approaches have inevitably been left outside this chapter's scope. Among them are those based on intelligent and fuzzy systems \cite{Delgado1998,Rodriguez2011,Herrera2014,Shengbao2017}. Additional insight into decision-making processes can also be obtained from various proxy systems of collective human behaviours, such as online social network systems \cite{Manning2023,Dankulov2015,Dankulov2022}, as well as through systematic studies of biosocial dynamics where multiscale approaches accounting for small-scale effects become essential \cite{Tadic2020,Tadic2021}. We will also mention the approaches based on inverse problem ideas and data-driven models, inverse Bayesian inference, inverse RL, and nonlinear programming \cite{Scales2000,Touqeer2022,Poulakakis2016,Gunji2021,Adrian2017}. Finally, the relevance of AI methodologies in our context has already been mentioned in Section \ref{sec:examplescollective}. Such methodologies can facilitate collective human decision-making and be invaluable in critical situations such as emergencies and rescue operations \cite{Wang2022Pre,Zhang2022,Whiten2022,Dwivedi2021,Kong2023,Hornischer2022}.
	 
	 \section{Remarks on human biosocial dynamics with complex psychological behaviour and nonequilibrium phenomena}
	
	What we discussed in the previous sections regarding biosocial human collective decision-making, linking neuroscience, mathematical modelling, and analysis (see Section 4), provides a good starting point for a deeper investigation of the role of nonequilibrium phenomena in biosocial and behavioural psychological dynamics and approaches. It is essential to realize that knowledge creation (see, e.g., \cite{Tadic2017} and references therein) and associated decision-making steps are nonequilibrium processes. Further, a critical element for decision-making is working memory, the brain's ability to store and recall information temporarily. In its turn, given that working memory is affected early during the onset of neurodegenerative diseases such as Alzheimer's, it can serve as a key to better understanding the course of such diseases and developing treatments.
	
	The next important remark is pertinent to the multiscale character of decision-making that we briefly mentioned in Section 5. We have in this research domain all the typical scales that we normally consider when dealing with nonequilibrium processes:
	\begin{itemize}
		\item { Microscopic scales:} strong energy fluctuations and complex information processing which, in their turn, induce a broad range of interactions, timescales, biochemical reactions, products and complexes, far from equilibrium considerations;
		\item { Kinetic / Mesoscopic scales:} here nonequilibrium processes may depart remarkably from standard scenarios of Gibbs ensemble averages and Boltzmann models, and interfacial transport and mixing provide prominent examples of nonequilibrium processes coupling kinetics to meso- and macroscopic scales;
		\item {Macroscopic scales:} with (multi) fields and (multiphase) flows considerations, we have interfacial mixing and nonequilibrium dynamics which can be (a) nonlocal (may include contributions from all the scales and may sense initial and boundary conditions \cite{Sytnyk2021}), (b) inhomogeneous (e.g., flow fields may not only be non-uniform, but they may involve fronts),
		(c) anisotropic (their dynamics depend on the directions), (d) statistically unsteady (with mean values of the quantities varying with time and with time-dependent fluctuations around these means), (e) and their invariance, correlations, and spectra may differ substantially from those of conventional fluid dynamics considerations, including turbulence, making corresponding CFD models frequently questionable in this field.
	\end{itemize}
	
However, with the above considerations, serious challenges quickly become apparent at the levels of theory, experiment, and fundamentals. While some have
common features with other nonequilibrium analyses, others also have their specifics for this research domain. At the level of theory, in the general setting, we have to deal with the multiscale, multiphase, nonlinear, nonlocal, and statistically unsteady character of the dynamics.
At the level of the experiment, interfacial mixing and nonequilibrium processes are a challenge to implement and study systematically in a well-controlled environment. Moreover, a systematic interpretation of these processes from the data is an additional challenge since the processes are statistically unsteady and may impose the influence of an observer on observational results.
At the fundamental level, interfacial mixing and nonequilibrium dynamics require a unified description of particles and fields across the scales based on a synergy of theory, simulations, and experiments. Successes in these areas open new opportunities for studying the fundamentals of interfaces and mixing and their nonequilibrium dynamics. Statistical mechanics considerations should not neglect its thermodynamic origin and the system's interaction with its environment. Hence, additional challenges come from the fact that in some environments, nonequilibrium dynamics of interfaces and (interfacial) mixing are expected to be enhanced. In contrast, in some others, such dynamics should be mitigated and/or tightly controlled.

When dealing with decision-making processes, including psychological behaviour into account, the drift-diffusion models play a vital role in the hierarchy of mathematical models, providing an initial stepping stone for moving to nonequilibrium phenomena. Recall, for example, that higher-level models such as the n-dimensional Fokker-Planck (FP) equation describe a drift-diffusion interplay with time-dependent drift (vector) and diffusion (matrix) coefficients for the probability density:
\begin{eqnarray}
	\frac{\partial P}{\partial t} = - \sum_{\mu=1}^{n} \frac{\partial}{\partial x_{\mu}} (v_{\mu} P) + \sum_{\mu, \nu=1}^{n} \frac{\partial}{\partial x_{\mu}} ( D_{\mu \nu} \frac{\partial P}{\partial x_{\nu}}).
\end{eqnarray}
It is well-known that one of many possible derivations of such models starts with a Markov process, with its continuous state described by a hierarchy of equations that collapses effectively just to the Chapman-Kolmogorov equation. Alternatively, we can start with the master equation - a Markov chain, where both the states and the time are discrete, considered in the limit where time is continuous, and then do a coarse-graining to get the FP equation via certain transformations such as the Kramers-Moyal expansion with its second-order truncation. It is also known that for a free energy functional such equations are equivalent to Markov processes defined on graphs (e.g., \cite{Chow2012}) and can be well-suited for brain network models, a connection we pursued in the previous two sections. Other motivational papers for our current consideration can also be mentioned (e.g.,
\cite{Jarzynski2008,Conti2022,Roldan2015,Wu2002,Vroylandt2022}). In analyzing social human collective decision-making, the DDM model is a prime tool in accounting for psychological features of biosocial dynamics. It is a model of sequential sampling that has been widely used in psychology and neuroscience to explain the observed patterns of choice and response times in a range
of binary-choice decision problems (e.g., \cite{Fudenberg2020}).


We described the main theoretical setting for the DDM in Section 2, and it is worthwhile also noting that it can be useful to carry out the Sequential Probability Ratio Test (SPRT) in that context, given that the SPRT plays a prominent role in psychological research (e.g.,  \cite{Stefan2022}). The simplest example could be to assume that the probability of seeing a measurement given the state is a Gaussian (normal) distribution where the mean ($\mu$) is different for the two states ($p(m_t| s = \pm 1)$) but the standard deviation ($\sigma$) is the same.  Then, by carrying out a series of measurements, we would like to figure out what the state is, given our measurements.  To do this, we can compare the total evidence up to time t (the final time of measurements) 
for our two hypotheses (that the state is +1 or that the state is -1). We can do this by computing a likelihood ratio, that is the ratio of the likelihood of all these measurements given the state is $+1, p(m_{1:t}|s = +1 )$  to the likelihood of the measurements given the state is $-1, p(m_{1:t}|s = -1 )$. This likelihood ratio test is typically quantified by taking the log of this likelihood ratio $\log \frac{p(m_t| s = +1)}{p(m_t| s = -1)}.$  Applying this framework, the influence of noise on decision-making can be analyzed.  Using the above-mentioned example and plotting the trajectories under the fixed stopping-time rule for different variances, one can easily conclude that a  higher noise would lead to the evidence accumulation varying up
and down more, translating this into the fact that humans are more likely to make a wrong decision with high noise. 
In this case, even if the actual distribution corresponds to $s = +1$, the accumulated log-likelihood ratio is more likely to be negative at the end. The situation is different when the variance is small because each new measurement will be very similar. Finally, in applications of this model to human decision-making, we also see that humans are more likely to be wrong with a small number of time steps before the decision, as there is more chance that the noise will affect the decision.  To move beyond such simple examples, we note that stochastic adaptation dynamics can be studied by using Hidden Markov Models (HMMs). Developing this idea into our context, we note that psychological behaviour can be studied through the associated adaptation dynamics of humans based on their sensory systems. To do that, in the modelling framework, the starting point can be taken as a Markov process where the joint distribution of observations and hidden states is modelled (that is,  both the prior distribution of hidden states, the transition probabilities, and the conditional distribution of observations given states, the emission probabilities). So, we come to a HMM, which is a generative type of models (because it is aimed at the joint probability distribution on a given observable variable and target variable). Then, the idea (see, e.g., \cite{Conti2022})  is to explore the connection between the adaptation dynamics, where the adaptation variable is driven out of equilibrium by an external stimulus, and the dissipation dynamics (measured by the rate of entropy production quantified as the relative entropy between forward and backward trajectories of the dynamics). Suppose we again assume that our state in the decision-making process can be either $+1$ or $-1$. In that case, the dynamics can be easily defined by the 2 x 2 matrix, and the equation for the evolution of the probability of the current state (represented by a two-dimensional vector) can easily be obtained.  Note that in this case, the elements of the matrix are defined by the probability of switching to state $s_t=j$ ($j = \pm 1$) from the previous state $s_{t-1} = i$ ($i = \pm 1$), that is by the conditional distribution $p(s_t = j|s_{t-1} = i)$. This HMM setting is inspired by neural brain systems. Indeed, we know that neural systems switch between discrete states, observable only indirectly, through their impact on neural activity. HMMs let us reason about these unobserved (hidden or latent) states using a time series of measurements. We are looking at the likelihood, which is the probability of our measurement $(m)$, given the hidden state $(s)$: $P(m|s)$. The immediate goal then would be to learn 
how changing the HMM’s transition probability and measurement noise could impact the data and how uncertainty increases as we predict the future (and, ultimately, how to gain information from the measurements).  In applying these ideas to the decision-making under uncertainties, our binary  variable $s_t \in \{-1,1\}$ become latent that switches randomly between the two states.
In the simplest setting, one can use a 1D Gaussian emission model $m_t|s_t \sim \mathcal{N}(\mu_{s_t}, \sigma_{s_t}^2)$ 
that provides evidence about the current state (see, e.g., \cite{Roweis1999,Bishop2007}).
Unlike the cases discussed earlier, now, in a HMM, we cannot directly observe the latent states $s_t$, because we get noisy measurements $m_t$ characterized by  $p(m|s_t)$. To move forward, we assume that humans have normal vision, colour recognition, auditory sense, and movement abilities. What is important to emphasize is that often humans have the wrong percept. In particular, they think their own route might be the best choice in escaping the emergency situations when the other neighbours might have other better choices of escape route; or vice versa. The  illusion in such situations is usually resolved once you gain a vision of the surroundings that lets you disambiguate the routes. However, the problem here has much deeper roots and the interested reader is encouraged to look at it from the thermodynamics point of view and decision-making in the arrow of time (e.g.,  \cite{Roldan2015,Gnesotto2020,Seif2021}). In our example, we use an HMM to model decision-making in the case of two alternative choices for choosing an escape route scenario. In particular, the state +1 represents the human choosing the nearest visible escape route, while state -1 stands for choosing another route.  While the 2 x 2 transition matrix in our HMM model can easily be constructed, simulations based on these considerations lead to some important conclusions regarding switching probabilistic state limits. This can be analyzed by plotting ``forgetting'' curves, the probability of switching states as a function of time. As the probability of switching increases (under the noise modelled by the Gaussian emission model that provides evidence about the current state), we eventually start observing oscillations, indicating that one forgets more quickly with high switching probability because the person becomes less certain that the state is the known one. 

The above considerations led us to believe that the adaptation should be cast as an inference problem for nonequilibrium dynamics so that we can use either Bayesian or Kalman filters for its solution. While we have emphasized the importance of inference problems throughout the previous sections, in the nonequilibrium context we refer the interested reader to (e.g., \cite{Conti2022}) for further details. In brief, the resulting filtered
adaptation dynamics couples the human (sensory) response function to
the state of the environment in a noisy manner, allowing for
their study in terms of stochastic (thermo)dynamics of nonequilibrium systems. It occurs due to the delay between changes in the stimulus's statistics and the adaptation mechanism's response, leading to irreversible
dynamics.  The latent state $s_t$ (more precisely, the location of the state $s$ at time $t$) will
evolve as a stochastic system in discrete time, with the evolution matrix satisfying certain conditions connected with nonequilibrium work relations (e.g.,  \cite{ Jarzynski2008}). For the reasons mentioned below, we use Kalman filtering in this case. A Kalman filter recursively estimates a posterior probability distribution over time using a mathematical model of the process and incoming measurements. This dynamic posterior allows us to improve our guess about the state's position; besides, its mean is the best estimate one can compute of the state’s actual position at each time step (see, e.g., \cite{Wu2002}). There are several open-source codes with algorithms for HMMs with Kalman filtering, among which we shall mention the open-source framework for Biological Neuron Models of the Neuromatch Academy. Using the associated algorithms,  we have analyzed and tracked how humans choose an escape route in emergency situations described in the previous section. In forward inference with HMM without filtering, we observe many outliers for the measurements compared with the true state (the measurements are considered as position over time according to the sensor). These are hidden states far away from the states +1 and -1, and such examples demonstrate that our estimations, in this case, are not really helpful in tracking how humans choose the nearest visible escape route or choose another route until they make a decision (for the simulations run to the time corresponding 100 s). In all these experiments, we assumed that humans will make a decision to choose the nearest visible escape route if the final value of the state is positive; otherwise, humans will choose another route with negative values. A very different situation has been obtained with the Kalman filter inference where in the case of non-equilibrium dynamics (modelled with continuous time Markov Chains), we saw that our estimations perfectly match the true state, showing that the HMM with Kalman filtering provides a viable tool to track how humans make their decisions. 

Our previous two sections studied the link between collective decision-making and brain networks. To bring this study to the next level, one should recall that brain dynamics are known to be nonequilibrium (e.g., \cite{Perl2021}). Until very recently, neural systems in general and brain network models, in particular, have been predominantly studied based on two main approaches: (a) biologically-inspired approaches (connectome-type of models), and (b) Bayesian approaches. While the latter approaches were also a part of our consideration here, it is worthwhile mentioning they were challenged empirically with experimental facts in decision-making, and the interested reader can find examples in the literature of the breakdown of the classical framework of cognitive science based on such approaches. While studying brain network models, we should resort to nonequilibrium information dynamics in the general setting. Indeed, it is well-known that the conditions for learning do not happen in an equilibrium state. In this case, whether physical or biological systems, they do have information systems features that can be studied via (a) information storage (memory), (b) information transfer (signalling), and (c) information modification (computation).  Hence, for the study of such information systems in the regime of nonequilibrium states, one cannot avoid challenges involving minimizing relative entropy (e.g., the Kullback-Leibler divergence) and looking at phase transitions near high mutual information with respect to different distributions. On a side note, this route should also be fundamental in clarifying 
Artificial Intelligence (AI) concepts when looking at nonequilibrium.  Our motivation in the nonequilibrium analysis of brain networks and associated decision-making is rooted in the analysis of neurodegenerative diseases (e.g., \cite{Pal2022,Shaheen2022,Pal2023}). Starting from earlier papers (e.g., \cite{Kensinger2003,Gagnon2011}), we know, for example, that Alzheimer's and other kinds of dementia affect working memory at an early stage. Hence, as a critical element in decision-making, working memory is also crucial for better understanding neurodegenerative diseases such as  Alzheimer’s disease and frontotemporal dementia (e.g., \cite{Stopford2012}).  For advanced approaches to model working memory, one can follow the premises where the FP equation is derived based on the Langevin equation, taking the latter as a starting point with a biophysical circuit model for working memory (e.g., \cite{Yan2020,Murray2017}) and apply one of the available methodologies for its solution such as the nonequilibrium landscape-flux method.


Our next remark concerns nonequilibrium dynamics, biosocial self-organization, and developmental psychology. Given that we are considering systems whose constituent elements consume energy, they are, by definition, out-of-equilibrium. They are active systems, and as a part of complex systems, their particularly attractive aspect is the emergence of cooperative phenomena, or self-organization, often driven by nonequilibrium dynamics  that relies on an external (energy) source. As we all know, paradigmatic examples here include flocking, collections of cells, etc, where the tools based on field theory, entropy production to measure to which degree the equilibrium is broken, and reaction-diffusion models can be applied. We have to deal with nonequilibrium processes coupling kinetics to meso- and macroscopic scales,  including  interfacial transport and mixing. The importance of microscopic scales further complicates the picture. This is coming not only from human biosocial dynamics, but also from their complex psychological behaviour, and they are closely interconnected. Despite that, these complex processes may still lead to self-organization and may thus expand opportunities for diagnostics and control of nonequilibrium dynamics. It should be emphasized that  the attempts to incorporate psychology with self-organization when dealing with biosocial dynamics of humans have been around at least since I. Prigogine got his Nobel Prize in 1977 (see, e.g., \cite{Brent1978,Chapman1991}, as well as later works in the context of human decision-making, e.g. \cite{West2012}, and others). Some of these and more recent works included renormalization group approaches for analyzing critical phenomena in complex systems with nonequilibrium features (e.g., \cite{Boettcher2011, Kaupuzs2022}),  as well as new ideas on evolving cycles and self-organized criticality in biosocial dynamics (see, e.g.,  \cite{Tadic2023}, and references therein).

Our final remark goes to nonequilibrium dynamics and AI methodologies. Considering AI tools and Machine Learning (ML) methods have emerged as an important direction to study problems in statistical mechanics, nonequilibrium phenomena pertinent to psychological behaviours and the biosocial dynamics of humans should not be an exception. Moreover, it is critical to do so. 
Indeed, we know that while the microscopic dynamics of physical systems are time reversible, the macroscopic world does not necessarily share this symmetry (e.g.,  \cite{Seif2021}). Further, as it was pointed out, fluctuations at small microscopic scales lead to an effective “blurring” of time’s arrow, and attempts have been made to quantify our ability to ``perceive'' its direction in a system-independent manner. In doing so, thermodynamic relationships such as the Clausius inequality can only be expressed in terms of averages, and such problems as the direction of time's arrow can be quantified as a problem in statistical inference in a way similar we discussed above. Further, we have already pointed out that since the general form of the nonequilibrium steady-state is not known (unlike the equilibrium case with the Boltzmann distribution),  generative models are an ideal candidate to model and learn these (nonequilibrium) distributions, with unsupervised learning techniques (whereas such nonequilibrium distributions as Kappa (e.g., \cite{Shizgal2018}) may serve as a testing ground for that). An example of such generative models based on HMMs has been discussed above.

Of course, while these ideas are critical in the areas we have discussed here, they can also be applied to such problems as the estimation of free energy differences, as well as to the identification of 
physical quantities that distinguish different regimes of
dynamics in out-of-equilibrium phenomena, but these considerations go beyond the scope of this chapter. Among future directions 
of the nonequilibrium considerations we considered in this section, we would like to mention (a) exploring a theoretical basis of the connection between the dynamics of quantum decision-making and the free-energy principle in cognitive science (e.g.,  \cite{Tanaka2022}), (b) refining mathematically concepts of nonequilibrium psychology (e.g., when a group of people agrees on something, and what they agreed upon is a consensus, and a consensus is by definition an equilibrium), and (c) expanding the area of applications where such concepts become decisive, including applications in nonequilibrium social science and policy; it requires  better estimates of policy consequences at the microlevel and advances in behavioural sciences revealing how people/firms/governments do behave in practice (e.g., \cite{Johnson2017}), and ultimately better predicting their behaviours.

	\section{Conclusions}\label{sec:Conclusions}
	
	We have proposed and described probabilistic drift-diffusion models and Bayesian inference frameworks for human social decision-making. In particular, we have provided details of the models and representative numerical examples. We also discussed the decision-making process in choosing an escape route scenario by considering the drift-diffusion models and Bayesian inference frameworks. Our numerical results have demonstrated that the right shortcut displays higher decoding accuracy than the wrong shortcut. Moreover, the average test accuracy is increased to 96.62\% even though the given data includes noise. The accuracy between the right shortcut and bad shortcut judgements in the case of a small window of data is lower than in the case of a full data set. 
	Furthermore, we have also provided a review of recent results on collective human decision-making and highlighted key approaches and challenges in analyzing human biosocial dynamics with complex psychological behaviour in the nonequilibrium setting. We examined recent developments in collective human decision-making and its applications with brain network models. We have found that neuromodulation and reinforcement learning methods are essential in human decision-making. Furthermore, the collective decision-making process has been scrutinized for cooperative and competitive groups, subject to different parameter choices. A better understanding of such decision-making systems would contribute to further developments of social human decision-making studies, higher-level brain-inspired functionality studies and other applications. 
	
	%
	\begin{acknowledgement}
		Authors are grateful to the NSERC and the CRC Program for their
		support. RM is also acknowledging support of the BERC 2022-2025 program and Spanish Ministry of Science, Innovation and Universities through the Agencia Estatal de Investigacion (AEI) BCAM Severo Ochoa excellence accreditation SEV-2017-0718.
	\end{acknowledgement}
	%
	%
   \bibliographystyle{ieeetr}
	\bibliography{mybibn.bib}

\begin{thebibliography}{100}

\bibitem{Fudenberg2020}
D.~Fudenberg, W.~Newey, P.~Strack, and T.~Strzalecki, ``Testing the
  drift-diffusion model,'' {\em PNAS}, vol.~117, no.~52, pp.~33141--33148,
  2020.

\bibitem{Vellmer2020}
S.~Vellmer and B.~Lindner, ``Decision-time statistics of nonlinear diffusion
  models: Characterizing long sequences of subsequent trials,'' {\em Journal of
  Mathematical Psychology}, vol.~99, p.~102445, 2020.

\bibitem{Smith2015}
P.~L. Smith, ``The poisson shot noise model of visual short-term memory and
  choice response time: Normalized coding by neural population size,'' {\em
  Journal of Mathematical Psychology}, vol.~66, pp.~41--52, 2015.

\bibitem{Mavrodiev2013}
C.~J.~T. P.~Mavrodiev and F.~Schweitzer, ``Quantifying the effects of social
  influence,'' {\em Scientific Reports}, vol.~3, p.~1360, 2013.

\bibitem{Mavrodiev2021}
P.~Mavrodiev and F.~Schweitzer, ``The ambigous role of social influence on the
  wisdom of crowds: An analytic approach,'' {\em Physica A}, vol.~567,
  p.~125624, 2021.

\bibitem{Haghani2019}
M.~Haghani and M.~Sarvi, ``Imitative (herd) behaviour in direction
  decision-making hinders efficiency of crowd evacuation processes,'' {\em
  Safety Science}, vol.~114, pp.~49--60, 2019.

\bibitem{Gold2007}
J.~I. Gold, M.~N. Shadlen, {\em et~al.}, ``The neural basis of decision
  making,'' {\em Annual Review of Neuroscience}, vol.~30, no.~1, pp.~535--574,
  2007.

\bibitem{Bogacz2006}
R.~Bogacz, E.~T. Brown, J.~Moehlis, P.~Holmes, and J.~D. Cohen, ``The physics
  of optimal decision making: a formal analysis of models of performance in
  two-alternative forced-choice tasks.,'' {\em Psychological Review},
  vol.~113(4), pp.~700--765, 2006.

\bibitem{Bitzer2014}
S.~Bitzer, H.~Park, F.~Blankenburg, and S.~J. Kiebel, ``Perceptual decision
  making: drift-diffusion model is equivalent to a bayesian model,'' {\em
  Frontiers in Human Neuroscience}, vol.~8, p.~102, 2014.

\bibitem{Fard2017}
P.~R. Fard, H.~Park, A.~Warkentin, S.~J. Kiebel, and S.~Bitzer, ``A bayesian
  reformulation of the extended drift-diffusion model in perceptual decision
  making,'' {\em Frontiers in Computational Neuroscience}, vol.~11, p.~29,
  2017.

\bibitem{Dokka2019}
K.~Dokka, H.~Park, M.~Jansen, G.~C. DeAngelis, and D.~E. Angelaki, ``Causal
  inference accounts for heading perception in the presence of object motion,''
  {\em Proceedings of the National Academy of Sciences}, vol.~116, no.~18,
  pp.~9060--9065, 2019.

\bibitem{Dokka2015}
K.~Dokka, G.~C. DeAngelis, and D.~E. Angelaki, ``Multisensory integration of
  visual and vestibular signals improves heading discrimination in the presence
  of a moving object,'' {\em Journal of Neuroscience}, vol.~35, no.~40,
  pp.~13599--13607, 2015.

\bibitem{Noel2022}
J.~Noel and D.~Angelaki, ``Cognitive, systems, and computational neurosciences
  of the self in motion,'' {\em Annual Review of Psychology}, vol.~73,
  pp.~103--129, 2022.

\bibitem{Ratcliff1978}
R.~Ratcliff, ``A theory of memory retrieval,'' {\em Psychological review},
  vol.~85, no.~2, p.~59, 1978.

\bibitem{Thieu2022coupled}
T.~K.~T. Thieu, A.~Muntean, and R.~Melnik, ``Coupled stochastic systems of
  {S}korokhod type: Well-posedness of a mathematical model and its
  applications,'' {\em Mathematical Methods in the Applied Sciences}, vol.~46,
  no.~6, pp.~7368--7390, 2023.

\bibitem{Pekkanen2022}
J.~Pekkanen, O.~T. Giles, Y.~M. Lee, R.~Madigan, T.~Daimon, N.~Merat, and
  G.~Markkula, ``Variable-drift diffusion models of pedestrian road-crossing
  decisions,'' {\em Computational Brain \& Behavior}, vol.~5, no.~1,
  pp.~60--80, 2022.

\bibitem{Melnik2000}
R.~Melnik and H.~He, ``Relaxation-time approximations of quasi-hydrodynamic
  type in semiconductor device modelling,'' {\em Modelling and Simulation in
  Materials Science and Engineering}, vol.~8, no.~2, p.~133, 2000.

\bibitem{Melnik2000quasi}
R.~Melnik and H.~He, ``Quasi-hydrodynamic modelling and computer simulation of
  coupled thermo-electrical processes in semiconductors,'' {\em Mathematics and
  computers in simulation}, vol.~52, no.~3-4, pp.~273--287, 2000.

\bibitem{Melnik2000modelling}
R.~Melnik and H.~He, ``Modelling nonlocal processes in semiconductor devices
  with exponential difference schemes,'' {\em Journal of Engineering
  Mathematics}, vol.~38, pp.~233--263, 2000.

\bibitem{Estrada2002}
R.~F. {Alvarez-Estrada}, ``{New hierarchy for the Liouville equation,
  irreversibility and Fokker‑Planck‑like structures},'' {\em Annalen der
  Physik}, vol.~514, pp.~357--385, May 2002.

\bibitem{Goddard2019}
B.~D. Goddard, T.~Hurst, and M.~Wilkinson, ``A derivation of the liouville
  equation for hard particle dynamics with non-conservative interactions,''
  {\em Proceedings of the Royal Society of Edinburgh: Section A Mathematics},
  vol.~151, pp.~1040 -- 1074, 2019.

\bibitem{Klein2022}
R.~Klein and L.~D. Site, ``Derivation of liouville-like equations for the
  n-state probability density of an open system with thermalized particle
  reservoirs and its link to molecular simulation,'' {\em Journal of Physics A:
  Mathematical and Theoretical}, vol.~55, p.~155002, mar 2022.

\bibitem{Degond2022}
P.~Degond, A.~Manhart, S.~Merino-Aceituno, D.~Peurichard, and L.~Sala, ``How
  environment affects active particle swarms: a case study,'' {\em Royal
  Society Open Science}, vol.~9, 2022.

\bibitem{Jiang2022}
Y.-Q. Jiang, Y.-G. Hu, and X.~Huang, ``Modeling pedestrian flow through a
  bottleneck based on a second-order continuum model,'' {\em Physica A:
  Statistical Mechanics and its Applications}, vol.~608, p.~128272, 2022.

\bibitem{Bellomo2022What}
N.~Bellomo, M.~Esfahanian, V.~Secchini, and P.~Terna, ``What is life? active
  particles tools towards behavioral dynamics in social-biology and
  economics,'' {\em Physics of Life Reviews}, vol.~43, pp.~189--207, 2022.

\bibitem{Bellomo2022towards}
N.~Bellomo, L.~Gibelli, A.~Quaini, and A.~Reali, ``Towards a mathematical
  theory of behavioral human crowds,'' {\em Mathematical Models and Methods in
  Applied Sciences}, vol.~32, no.~02, pp.~321--358, 2022.

\bibitem{Ratcliff2016}
R.~Ratcliff, P.~L. Smith, S.~D. Brown, and G.~McKoon, ``Diffusion decision
  model: Current issues and history,'' {\em Trends in Cognitive Sciences},
  vol.~20, no.~4, pp.~260--281, 2016.

\bibitem{Zhu2023}
S.~I. Zhu and G.~J. Goodhill, ``From perception to behavior: The neural
  circuits underlying prey hunting in larval zebrafish,'' {\em Frontiers in
  Neural Circuits}, vol.~17, 2023.

\bibitem{Saraiva2023}
T.~Saraiva and T.~C. Gonçalves, ``The role of emotions and knowledge on
  preference for uncertainty: Follow your heart but listen to your brain!,''
  {\em Risks}, vol.~11, no.~1, 2023.

\bibitem{Castagna2023}
P.~J. Castagna, S.~{van Noordt}, P.~B. Sederberg, and M.~J. Crowley, ``Modeling
  brain dynamics and gaze behavior: Starting point bias and drift rate relate
  to frontal midline theta oscillations,'' {\em NeuroImage}, vol.~268,
  p.~119871, 2023.

\bibitem{Myers2022}
C.~E. Myers, A.~Interian, and A.~A. Moustafa, ``A practical introduction to
  using the drift diffusion model of decision-making in cognitive psychology,
  neuroscience, and health sciences,'' {\em Frontiers in Psychology}, vol.~13,
  2022.

\bibitem{Wiecki2013}
T.~V. Wiecki, I.~Sofer, and M.~J. Frank, ``Hddm: Hierarchical bayesian
  estimation of the drift-diffusion model in python,'' {\em Frontiers in
  Neuroinformatics}, p.~14, 2013.

\bibitem{Pedersen2017}
M.~L. Pedersen, M.~J. Frank, and G.~Biele, ``The drift diffusion model as the
  choice rule in reinforcement learning,'' {\em Psychonomic Bulletin \&
  Review}, vol.~24, pp.~1234--1251, 2017.

\bibitem{Boelts2022}
J.~Boelts, J.-M. Lueckmann, R.~Gao, and J.~H. Macke, ``Flexible and efficient
  simulation-based inference for models of decision-making,'' {\em Elife},
  vol.~11, p.~e77220, 2022.

\bibitem{Thapa2022}
S.~Thapa, S.~Park, Y.~Kim, J.-H. Jeon, R.~Metzler, and M.~A. Lomholt,
  ``Bayesian inference of scaled versus fractional brownian motion,'' {\em
  Journal of Physics A: Mathematical and Theoretical}, vol.~55, no.~19,
  p.~194003, 2022.

\bibitem{Fengler2022}
A.~Fengler, K.~Bera, M.~L. Pedersen, and M.~J. Frank, ``Beyond drift diffusion
  models: Fitting a broad class of decision and reinforcement learning models
  with hddm,'' {\em Journal of Cognitive Neuroscience}, vol.~34, no.~10,
  pp.~1780--1805, 2022.

\bibitem{Manning2023}
T.~S. Manning, B.~N. Naecker, I.~R. McLean, B.~Rokers, J.~W. Pillow, and E.~A.
  Cooper, ``A general framework for inferring bayesian ideal observer models
  from psychophysical data,'' {\em eNeuro}, vol.~10, no.~1, 2023.

\bibitem{Tadic2017}
B.~Tadi{\'c}, M.~M. Dankulov, and R.~Melnik, ``Mechanisms of self-organized
  criticality in social processes of knowledge creation,'' {\em Physical Review
  E}, vol.~96, no.~3, p.~032307, 2017.

\bibitem{Tadic2021Self}
B.~Tadić and R.~Melnik, ``Self-organised critical dynamics as a key to
  fundamental features of complexity in physical, biological, and social
  networks,'' {\em Dynamics}, vol.~1, no.~2, pp.~181--197, 2021.

\bibitem{Ramstead2020}
M.~J. Ramstead, M.~D. Kirchhoff, and K.~J. Friston, ``A tale of two densities:
  active inference is enactive inference,'' {\em Adaptive Behavior}, vol.~28,
  no.~4, pp.~225--239, 2020.

\bibitem{Isomura2022}
T.~Isomura, ``Active inference leads to bayesian neurophysiology,'' {\em
  Neuroscience Research}, vol.~175, pp.~38--45, 2022.

\bibitem{Costa2022}
L.~Da~Costa, P.~Lanillos, N.~Sajid, K.~Friston, and S.~Khan, ``How active
  inference could help revolutionise robotics,'' {\em Entropy}, vol.~24, no.~3,
  2022.

\bibitem{Badcock2022}
P.~B. {Badcock}, M.~J.~D. {Ramstead}, Z.~{Sheikhbahaee}, and A.~{Constant},
  ``{Applying the Free Energy Principle to Complex Adaptive Systems},'' {\em
  Entropy}, vol.~24, p.~689, May 2022.

\bibitem{Kiverstein2022}
J.~Kiverstein, M.~D. Kirchhoff, and T.~Froese, ``The problem of meaning: The
  free energy principle and artificial agency,'' {\em Frontiers in
  Neurorobotics}, vol.~16, 2022.

\bibitem{Xu2020}
T.~Xu, D.~Shi, J.~Chen, T.~Li, P.~Lin, and J.~Ma, ``Dynamics of emotional
  contagion in dense pedestrian crowds,'' {\em Physics Letters A}, vol.~384,
  no.~3, p.~126080, 2020.

\bibitem{Iinuma2021}
K.~Iinuma and K.~Kogiso, ``Emotion-involved human decision-making model,'' {\em
  Mathematical and Computer Modelling of Dynamical Systems}, vol.~27, no.~1,
  pp.~543--561, 2021.

\bibitem{Fu2021}
M.~Fu, R.~Liu, and Y.~Zhang, ``Why do people make risky decisions during a fire
  evacuation? study on the effect of smoke level, individual risk preference,
  and neighbor behavior,'' {\em Safety Science}, vol.~140, p.~105245, 2021.

\bibitem{Hoffmann2013}
A.~O. Hoffmann, S.~F. Henry, and N.~Kalogeras, ``Aspirations as reference
  points: an experimental investigation of risk behavior over time,'' {\em
  Theory and Decision}, vol.~75, pp.~193--210, 2013.

\bibitem{Kwon2022}
J.-H. Kwon, J.~Kim, S.~Kim, and G.-H. Cho, ``Pedestrians safety perception and
  crossing behaviors in narrow urban streets: An experimental study using
  immersive virtual reality technology,'' {\em Accident Analysis \&
  Prevention}, vol.~174, p.~106757, 2022.

\bibitem{Usher2001}
M.~Usher and J.~L. McClelland, ``The time course of perceptual choice: the
  leaky, competing accumulator model,'' {\em Psychological Review}, vol.~108 3,
  pp.~550--92, 2001.

\bibitem{Roy2021}
N.~A. Roy, J.~H. Bak, A.~Akrami, C.~D. Brody, and J.~W. Pillow, ``Extracting
  the dynamics of behavior in sensory decision-making experiments,'' {\em
  Neuron}, vol.~109, no.~4, pp.~597--610, 2021.

\bibitem{Singer2018}
Y.~Singer, Y.~Teramoto, B.~D. Willmore, J.~W. Schnupp, A.~J. King, and N.~S.
  Harper, ``Sensory cortex is optimized for prediction of future input,'' {\em
  eLife}, vol.~7, p.~e31557, jun 2018.

\bibitem{Miller2013}
P.~Miller, {\em Decision Making, Threshold}, pp.~1--4.
\newblock New York, NY: Springer New York, 2013.

\bibitem{Mormann2010}
M.~Mormann, J.~Malmaud, A.~G. Huth, C.~Koch, and A.~Rangel, ``The drift
  diffusion model can account for the accuracy and reaction time of value-based
  choices under high and low time pressure,'' in {\em Judgment and Decision
  Making}, 2010.

\bibitem{Masis2023}
J.~Masís, T.~Chapman, J.~Y. Rhee, D.~D. Cox, and A.~M. Saxe, ``Strategically
  managing learning during perceptual decision making,'' {\em eLife}, vol.~12,
  p.~e64978, 2023.

\bibitem{Gerwinn2007}
S.~Gerwinn, M.~Bethge, J.~H. Macke, and M.~Seeger, ``Bayesian inference for
  spiking neuron models with a sparsity prior,'' {\em Advances in Neural
  Information Processing Systems}, vol.~20, 2007.

\bibitem{Macke2011}
J.~H. Macke, L.~Buesing, J.~P. Cunningham, B.~M. Yu, K.~V. Shenoy, and
  M.~Sahani, ``Empirical models of spiking in neural populations,'' {\em
  Advances in neural information processing systems}, vol.~24, 2011.

\bibitem{Kleinbaum2002}
D.~G. Kleinbaum, K.~Dietz, M.~Gail, M.~Klein, and M.~Klein, {\em Logistic
  Regression}.
\newblock Springer, 2002.

\bibitem{Weber2017}
A.~I. Weber and J.~W. Pillow, ``Capturing the dynamical repertoire of single
  neurons with generalized linear models,'' {\em Neural Computation}, vol.~29,
  no.~12, pp.~3260--3289, 2017.

\bibitem{Theis2013}
L.~Theis, A.~M. Chagas, D.~Arnstein, C.~Schwarz, and M.~Bethge, ``Beyond glms:
  a generative mixture modeling approach to neural system identification,''
  {\em PLoS computational biology}, vol.~9, no.~11, p.~e1003356, 2013.

\bibitem{Fortuna2023}
L.~Fortuna and A.~Buscarino, ``Spiking neuron mathematical models: A compact
  overview,'' {\em Bioengineering}, vol.~10, no.~2, p.~174, 2023.

\bibitem{Anumula2018}
J.~Anumula, D.~Neil, T.~Delbruck, and S.-C. Liu, ``Feature representations for
  neuromorphic audio spike streams,'' {\em Frontiers in Neuroscience}, vol.~12,
  p.~23, 2018.

\bibitem{Wang2021neural}
Q.~Wang, Y.~Wang, P.~Wang, M.~Peng, M.~Zhang, Y.~Zhu, S.~Wei, C.~Chen, X.~Chen,
  S.~Luo, {\em et~al.}, ``Neural representations of the amount and the delay
  time of reward in intertemporal decision making,'' {\em Human Brain Mapping},
  vol.~42, no.~11, pp.~3450--3469, 2021.

\bibitem{Cerulli2022}
G.~Cerulli, ``Machine learning using stata/python,'' {\em The Stata Journal},
  vol.~22, no.~4, pp.~772--810, 2022.

\bibitem{Nelli2023}
S.~Nelli, L.~Braun, T.~Dumbalska, A.~Saxe, and C.~Summerfield, ``Neural
  knowledge assembly in humans and neural networks,'' {\em Neuron}, 2023.

\bibitem{Reina2021}
A.~Reina, E.~Ferrante, and G.~Valentini, ``Collective decision-making in living
  and artificial systems,'' {\em Swarm Intelligence}, vol.~15, no.~1, pp.~1--6,
  2021.

\bibitem{Bose2017}
T.~Bose, A.~Reina, and J.~A. Marshall, ``Collective decision-making,'' {\em
  Current Opinion in Behavioral Sciences}, vol.~16, pp.~30--34, 2017.

\bibitem{Teodorescu2016}
A.~Teodorescu, R.~Moran, and M.~Usher, ``Absolutely relative or relatively
  absolute: violations of value invariance in human decision making,'' {\em
  Psychonomic Bulletin \& Review}, vol.~23, pp.~22--38, 2016.

\bibitem{Galam2008}
S.~Galam, ``Sociophysics: A review of galam models,'' {\em International
  Journal of Modern Physics C}, vol.~19, no.~03, pp.~409--440, 2008.

\bibitem{Molavi2018}
P.~Molavi, A.~Tahbaz-Salehi, and A.~Jadbabaie, ``A theory of non-bayesian
  social learning,'' {\em Econometrica}, vol.~86, no.~2, pp.~445--490, 2018.

\bibitem{Groeber2014}
J.~L. P.~Groeber and F.~Schweitzer, ``Dissonance minimization as a
  microfoundation of social influence in models of opinion formation,'' {\em
  Journal of Mathematical Sociology}, vol.~38, pp.~147--174, 2014.

\bibitem{Rauhut2011}
H.~Rauhut and J.~Lorenz, ``The wisdom of crowds in one mind: How individuals
  can simulate the knowledge of diverse societies to reach better decisions,''
  {\em Journal of Mathematical Psychology}, vol.~55, pp.~191--197, 2011.

\bibitem{Barfuss2022}
W.~Barfuss, ``Dynamical systems as a level of cognitive analysis of multi-agent
  learning: Algorithmic foundations of temporal-difference learning dynamics,''
  {\em Neural Computing and Applications}, vol.~34, no.~3, pp.~1653--1671,
  2022.

\bibitem{Adler2002}
J.~L. Adler and V.~J. Blue, ``A cooperative multi-agent transportation
  management and route guidance system,'' {\em Transportation Research Part C:
  Emerging Technologies}, vol.~10, no.~5-6, pp.~433--454, 2002.

\bibitem{Melnik2009coupling}
R.~V. Melnik, ``Coupling control and human factors in mathematical models of
  complex systems,'' {\em Engineering Applications of Artificial Intelligence},
  vol.~22, no.~3, pp.~351--362, 2009.

\bibitem{Bulling2014}
N.~Bulling, ``A survey of multi-agent decision making,'' {\em KI-K{\"u}nstliche
  Intelligenz}, vol.~28, pp.~147--158, 2014.

\bibitem{Geng2020}
B.~Geng, S.~Brahma, T.~Wimalajeewa, P.~K. Varshney, and M.~Rangaswamy,
  ``Prospect theoretic utility based human decision making in multi-agent
  systems,'' {\em IEEE Transactions on Signal Processing}, vol.~68,
  pp.~1091--1104, 2020.

\bibitem{Tadic2020}
B.~Tadi\'{c} and R.~Melnik, ``Modeling latent infection transmissions through
  biosocial stochastic dynamics,'' {\em PLoS ONE}, vol.~15, no.~10,
  p.~e0241163, 2020.

\bibitem{Tadic2021}
{B. Tadi\'{c} and R. Melnik}, ``Microscopic dynamics modeling unravels the role
  of asymptomatic virus carriers in {S}{A}{R}{S}-{C}o{V}-2 epidemics at the
  interplay between biological and social factors,'' {\em Computers in Biology
  and Medicine}, vol.~133, p.~104422, 2021.

\bibitem{Schweitzer2020}
F.~Schweitzer, T.~Krivachy, and D.~Garcia, ``An agent-based model of opinion
  polarization driven by emotions,'' {\em Complexity}, vol.~2020, pp.~1--11,
  2020.

\bibitem{Bose2018}
T.~Bose, A.~Reina, and J.~A.~R. Marshall, ``Inhibition and excitation shape
  activity selection: effect of oscillations in a decision-making circuit,''
  {\em Neural Computation}, vol.~31, no.~5, pp.~870--896, 2019.

\bibitem{Park2019}
S.~A. Park, M.~Sestito, E.~D. Boorman, and J.-C. Dreher, ``Neural computations
  underlying strategic social decision-making in groups,'' {\em Nature
  Communications}, vol.~10, no.~1, p.~5287, 2019.

\bibitem{Reverberi2022}
C.~Reverberi, D.~Pischedda, M.~Mantovani, J.-D. Haynes, and A.~Rustichini,
  ``Strategic complexity and cognitive skills affect brain response in
  interactive decision-making,'' {\em Scientific Reports}, vol.~12, no.~1,
  p.~15896, 2022.

\bibitem{Vijayakumar2011}
S.~Vijayakumar, T.~Hospedales, and A.~Haith, ``Generative probabilistic
  modeling: understanding causal sensorimotor integration,'' {\em Sensory Cue
  Integration}, pp.~63--81, 2011.

\bibitem{Doya2007}
K.~Doya, S.~Ishii, A.~Pouget, and R.~P. Rao, {\em Bayesian brain: Probabilistic
  approaches to neural coding}.
\newblock MIT press, 2007.

\bibitem{Acerbi2014}
L.~Acerbi, W.~J. Ma, and S.~Vijayakumar, ``A framework for testing
  identifiability of bayesian models of perception,'' {\em Advances in neural
  information processing systems}, vol.~27, 2014.

\bibitem{Ma2023}
W.~J. Ma, K.~P. Kording, and D.~Goldreich, {\em Bayesian Models of Perception
  and Action: An Introduction}.
\newblock MIT press, 2023.

\bibitem{Friston2010}
K.~Friston, ``The free-energy principle: a unified brain theory?,'' {\em Nature
  Reviews Neuroscience}, vol.~11, no.~2, pp.~127--138, 2010.

\bibitem{Parr2019}
T.~Parr and K.~J. Friston, ``Generalised free energy and active inference,''
  {\em Biological Cybernetics}, vol.~113, pp.~495 -- 513, 2019.

\bibitem{Khalvati2019}
K.~Khalvati, S.~A. Park, S.~Mirbagheri, R.~Philippe, M.~Sestito, J.-C. Dreher,
  and R.~P. Rao, ``Modeling other minds: Bayesian inference explains human
  choices in group decision-making,'' {\em Science Advances}, vol.~5, no.~11,
  p.~eaax8783, 2019.

\bibitem{Zhang2020}
L.~Zhang and J.~Gl{\"a}scher, ``A brain network supporting social influences in
  human decision-making,'' {\em Science Advances}, vol.~6, no.~34, p.~eabb4159,
  2020.

\bibitem{Kumar2022}
S.~Steixner-Kumar, T.~Rusch, P.~Doshi, M.~Spezio, and J.~Gläscher, ``Humans
  depart from optimal computational models of interactive decision-making
  during competition under partial information,'' {\em Scientific Reports},
  vol.~12, 01 2022.

\bibitem{Tarcai2011}
N.~Tarcai, C.~Vir{\'a}gh, D.~{\'A}bel, M.~Nagy, P.~L. V{\'a}rkonyi,
  G.~V{\'a}s{\'a}rhelyi, and T.~Vicsek, ``Patterns, transitions and the role of
  leaders in the collective dynamics of a simple robotic flock,'' {\em Journal
  of Statistical Mechanics: Theory and Experiment}, vol.~2011, no.~04,
  p.~P04010, 2011.

\bibitem{Wang2022}
R.~Wang, F.~Fang, J.~Cui, and W.~Zheng, ``Learning self-driven collective
  dynamics with graph networks,'' {\em Scientific Reports}, vol.~12, no.~1,
  pp.~1--11, 2022.

\bibitem{Valentini2014}
G.~Valentini, H.~Hamann, and M.~Dorigo, ``Self-organized collective decision
  making: The weighted voter model,'' vol.~1, 01 2014.

\bibitem{Hesse2014}
J.~Hesse and T.~Gross, ``Self-organized criticality as a fundamental property
  of neural systems,'' {\em Frontiers in Systems Neuroscience}, vol.~8, p.~166,
  2014.

\bibitem{De2006}
L.~De~Arcangelis, C.~Perrone-Capano, and H.~J. Herrmann, ``Self-organized
  criticality model for brain plasticity,'' {\em Physical Review Letters},
  vol.~96, no.~2, p.~028107, 2006.

\bibitem{Plenz2021}
D.~Plenz, T.~L. Ribeiro, S.~R. Miller, P.~A. Kells, A.~Vakili, and E.~L. Capek,
  ``Self-organized criticality in the brain,'' {\em Frontiers in Physics},
  vol.~9, p.~639389, 2021.

\bibitem{Zhang2021}
Z.~Zhang and L.~Jia, ``Optimal guidance strategy for crowd evacuation with
  multiple exits: A hybrid multiscale modeling approach,'' {\em Applied
  Mathematical Modelling}, vol.~90, pp.~488--504, 2021.

\bibitem{Boehm2021}
U.~Boehm, S.~Cox, G.~Gantner, and R.~Stevenson, ``Fast solutions for the
  first-passage distribution of diffusion models with space-time-dependent
  drift functions and time-dependent boundaries,'' {\em Journal of Mathematical
  Psychology}, vol.~105, p.~102613, 2021.

\bibitem{Boehm2022}
U.~Boehm, S.~Cox, G.~Gantner, and R.~Stevenson, ``Efficient numerical
  approximation of a non-regular {F}okker--{P}lanck equation associated with
  first-passage time distributions,'' {\em BIT Numerical Mathematics}, vol.~62,
  p.~1355–1382, 2022.

\bibitem{Sander2020}
D.~Sander, ``Comment: Collective epistemic emotions and individualized
  learning: A relational account,'' {\em Emotion Review}, vol.~12, no.~4,
  pp.~230--232, 2020.

\bibitem{Brown2021}
C.~L. Brown, K.-H. Chen, J.~L. Wells, M.~C. Otero, D.~E. Connelly, R.~W.
  Levenson, and B.~L. Fredrickson, ``Shared emotions in shared lives: Moments
  of co-experienced affect, more than individually experienced affect, linked
  to relationship quality.,'' {\em Emotion}, 2021.

\bibitem{Goldenberg2020}
A.~Goldenberg, D.~Garcia, E.~Halperin, and J.~J. Gross, ``Collective
  emotions,'' {\em Current Directions in Psychological Science}, vol.~29,
  no.~2, pp.~154--160, 2020.

\bibitem{Metzler2022}
H.~Metzler, B.~Rim{\'e}, M.~Pellert, T.~Niederkrotenthaler, A.~Di~Natale, and
  D.~Garcia, ``Collective emotions during the covid-19 outbreak.,'' {\em
  Emotion}, 2022.

\bibitem{Bogacz2007TheBG}
R.~Bogacz and K.~N. Gurney, ``The basal ganglia and cortex implement optimal
  decision making between alternative actions,'' {\em Neural Computation},
  vol.~19, pp.~442--477, 2007.

\bibitem{Bang2020}
D.~Bang, K.~T. Kishida, T.~Lohrenz, J.~P. White, A.~W. Laxton, S.~B. Tatter,
  S.~M. Fleming, and P.~R. Montague, ``Sub-second dopamine and serotonin
  signaling in human striatum during perceptual decision-making,'' {\em
  Neuron}, vol.~108, no.~5, pp.~999--1010, 2020.

\bibitem{Marzecova2021}
A.~Marzecov{\'a}, L.~F. Kaiser, and A.~Maddah, ``Neuromodulation of foraging
  decisions: The role of dopamine,'' {\em Frontiers in Behavioral
  Neuroscience}, vol.~15, p.~660667, 2021.

\bibitem{Grossman2022}
C.~D. Grossman and J.~Y. Cohen, ``Neuromodulation and neurophysiology on the
  timescale of learning and decision-making,'' {\em Annual Review of
  Neuroscience}, vol.~45, no.~1, pp.~317--337, 2022.

\bibitem{Montague2004}
P.~R. Montague, S.~E. Hyman, and J.~D. Cohen, ``Computational roles for
  dopamine in behavioural control,'' {\em Nature}, vol.~431, no.~7010,
  pp.~760--767, 2004.

\bibitem{Iglesias2021}
S.~Iglesias, L.~Kasper, S.~J. Harrison, R.~Manka, C.~Mathys, and K.~E. Stephan,
  ``Cholinergic and dopaminergic effects on prediction error and uncertainty
  responses during sensory associative learning,'' {\em NeuroImage}, vol.~226,
  p.~117590, 2021.

\bibitem{Sojitra2018}
R.~B. Sojitra, I.~Lerner, J.~R. Petok, and M.~A. Gluck, ``Age affects
  reinforcement learning through dopamine-based learning imbalance and high
  decision noise—not through parkinsonian mechanisms,'' {\em Neurobiology of
  Aging}, vol.~68, pp.~102--113, 2018.

\bibitem{Oettl2020}
L.-L. Oettl, M.~Scheller, C.~Filosa, S.~Wieland, F.~Haag, C.~Loeb,
  D.~Durstewitz, R.~Shusterman, E.~Russo, and W.~Kelsch, ``Phasic dopamine
  reinforces distinct striatal stimulus encoding in the olfactory tubercle
  driving dopaminergic reward prediction,'' {\em Nature Communications},
  vol.~11, no.~1, p.~3460, 2020.

\bibitem{Belkaid2020}
M.~Belkaid and J.~L. Krichmar, ``Modeling uncertainty-seeking behavior mediated
  by cholinergic influence on dopamine,'' {\em Neural Networks}, vol.~125,
  pp.~10--18, 2020.

\bibitem{Botvinick2012}
M.~M. Botvinick, ``Hierarchical reinforcement learning and decision making,''
  {\em Current Opinion in Neurobiology}, vol.~22, no.~6, pp.~956--962, 2012.

\bibitem{Dabney2020}
W.~Dabney, Z.~Kurth-Nelson, N.~Uchida, C.~K. Starkweather, D.~Hassabis,
  R.~Munos, and M.~M. Botvinick, ``A distributional code for value in
  dopamine-based reinforcement learning,'' {\em Nature}, vol.~577, pp.~671 --
  675, 2020.

\bibitem{Rmus2021}
M.~Rmus, S.~D. McDougle, and A.~G. Collins, ``The role of executive function in
  shaping reinforcement learning,'' {\em Current Opinion in Behavioral
  Sciences}, vol.~38, pp.~66--73, 2021.

\bibitem{Eckstein2021}
M.~K. Eckstein, L.~Wilbrecht, and A.~G. Collins, ``What do reinforcement
  learning models measure? interpreting model parameters in cognition and
  neuroscience,'' {\em Current Opinion in Behavioral Sciences}, vol.~41,
  pp.~128--137, 2021.

\bibitem{Lin2022}
B.~Lin, D.~Bouneffouf, and G.~Cecchi, ``Predicting human decision making in
  psychological tasks with recurrent neural networks,'' {\em PLOS ONE},
  vol.~17, pp.~1--18, 05 2022.

\bibitem{Hassabis2017}
D.~Hassabis, D.~Kumaran, C.~Summerfield, and M.~Botvinick,
  ``Neuroscience-inspired artificial intelligence,'' {\em Neuron}, vol.~95,
  no.~2, pp.~245--258, 2017.

\bibitem{Chen2022}
Y.~Chen, Z.~Wei, H.~Gou, H.~Liu, L.~Gao, X.~He, and X.~Zhang, ``How far is
  brain-inspired artificial intelligence away from brain?,'' {\em Frontiers in
  Neuroscience}, vol.~16, 2022.

\bibitem{Sutton2018}
R.~S. Sutton and A.~G. Barto, {\em Reinforcement learning: An introduction}.
\newblock MIT press, 2018.

\bibitem{Faisal2008}
A.~A. Faisal, L.~P. Selen, and D.~M. Wolpert, ``Noise in the nervous system,''
  {\em Nature Reviews Neuroscience}, vol.~9, no.~4, pp.~292--303, 2008.

\bibitem{Gallup2012}
A.~C. Gallup, J.~J. Hale, D.~J.~T. Sumpter, S.~Garnier, A.~Kacelnik, J.~R.
  Krebs, and I.~D. Couzin, ``Visual attention and the acquisition of
  information in human crowds,'' {\em Proceedings of the National Academy of
  Sciences}, vol.~109, no.~19, pp.~7245--7250, 2012.

\bibitem{Bragge2017}
J.~Bragge, H.~Kallio, T.~Sepp{\"a}l{\"a}, T.~Lainema, and P.~Malo,
  ``Decision-making in a real-time business simulation game: cultural and
  demographic aspects in small group dynamics,'' {\em International Journal of
  Information Technology \& Decision Making}, vol.~16, no.~03, pp.~779--815,
  2017.

\bibitem{Tump2022}
A.~N. Tump, M.~Wolf, P.~Romanczuk, and R.~H. J.~M. Kurvers, ``Avoiding costly
  mistakes in groups: The evolution of error management in collective decision
  making,'' {\em PLOS Computational Biology}, vol.~18, no.~8, pp.~1--21, 2022.

\bibitem{Hamblin2013}
S.~Hamblin, ``On the practical usage of genetic algorithms in ecology and
  evolution,'' {\em Methods in Ecology and Evolution}, vol.~4, no.~2,
  pp.~184--194, 2013.

\bibitem{Delgado1998}
M.~Delgado, F.~Herrera, E.~Herrera-Viedma, and L.~Martinez, ``Combining
  numerical and linguistic information in group decision making,'' {\em
  Information Sciences}, vol.~107, no.~1-4, pp.~177--194, 1998.

\bibitem{Rodriguez2011}
R.~M. Rodriguez, L.~Martinez, and F.~Herrera, ``Hesitant fuzzy linguistic term
  sets for decision making,'' {\em IEEE Transactions on Fuzzy Systems},
  vol.~20, no.~1, pp.~109--119, 2011.

\bibitem{Herrera2014}
F.~Herrera, L.~Mart{\'\i}nez-L{\'o}pez, V.~Torra, and Z.~Xu, ``Hesitant fuzzy
  sets: An emerging tool in decision making.,'' {\em Int. J. Intell. Syst.},
  vol.~29, no.~6, pp.~493--494, 2014.

\bibitem{Shengbao2017}
S.~Yao and J.~Hu, ``Combining comparative linguistic expressions and numerical
  information in multi-attribute group decision making—a simulation-based
  approach,'' {\em J. Intell. Fuzzy Syst.}, vol.~33, no.~6, p.~3835–3852,
  2017.

\bibitem{Dankulov2015}
M.~M. Dankulov, R.~Melnik, and B.~Tadi{\'c}, ``The dynamics of meaningful
  social interactions and the emergence of collective knowledge,'' {\em
  Scientific Reports}, vol.~5, no.~1, p.~12197, 2015.

\bibitem{Dankulov2022}
M.~Mitrović~Dankulov, B.~Tadić, and R.~Melnik, ``Analysis of worldwide
  time-series data reveals some universal patterns of evolution of the
  sars-cov-2 pandemic,'' {\em Frontiers in Physics}, vol.~10, 2022.

\bibitem{Scales2000}
J.~A. Scales and R.~Snieder, ``The anatomy of inverse problems,'' {\em
  GEOPHYSICS}, vol.~65, no.~6, pp.~1708--1710, 2000.

\bibitem{Touqeer2022}
M.~Touqeer, R.~Umer, A.~Ahmadian, S.~Salahshour, and M.~Salimi, ``Signed
  distance-based closeness coefficients approach for solving inverse non-linear
  programming models for multiple criteria group decision-making using interval
  type-2 pythagorean fuzzy numbers,'' {\em Granular Computing}, vol.~7,
  pp.~881--901, 11 2021.

\bibitem{Poulakakis2016}
I.~Poulakakis, G.~F. Young, L.~Scardovi, and N.~E. Leonard, ``Information
  centrality and ordering of nodes for accuracy in noisy decision-making
  networks,'' {\em IEEE Transactions on Automatic Control}, vol.~61, no.~4,
  pp.~1040--1045, 2016.

\bibitem{Gunji2021}
Y.-P. Gunji, T.~Kawai, H.~Murakami, T.~Tomaru, M.~Minoura, and S.~Shinohara,
  ``Lévy walk in swarm models based on bayesian and inverse bayesian
  inference,'' {\em Computational and Structural Biotechnology Journal},
  vol.~19, pp.~247--260, 2021.

\bibitem{Adrian2017}
A.~\v{S}o\v{s}i\'{c}, W.~R. KhudaBukhsh, A.~M. Zoubir, and H.~Koeppl, ``Inverse
  reinforcement learning in swarm systems,'' in {\em Proceedings of the 16th
  Conference on Autonomous Agents and MultiAgent Systems}, p.~1413–1421,
  International Foundation for Autonomous Agents and Multiagent Systems, 2017.

\bibitem{Wang2022Pre}
Z.~Wang, T.~Zhang, X.~Wu, and X.~Huang, ``Predicting transient building fire
  based on external smoke images and deep learning,'' {\em Journal of Building
  Engineering}, vol.~47, p.~103823, 2022.

\bibitem{Zhang2022}
T.~Zhang, Z.~Wang, H.~Y. Wong, W.~C. Tam, X.~Huang, and F.~Xiao, ``Real-time
  forecast of compartment fire and flashover based on deep learning,'' {\em
  Fire Safety Journal}, vol.~130, p.~103579, 2022.

\bibitem{Whiten2022}
A.~Whiten, D.~Biro, N.~Bredeche, E.~Garland, and S.~Kirby, ``The emergence of
  collective knowledge and cumulative culture in animals, humans and
  machines,'' vol.~377, 01 2022.

\bibitem{Dwivedi2021}
Y.~K. Dwivedi, L.~Hughes, E.~Ismagilova, G.~Aarts, C.~Coombs, T.~Crick,
  Y.~Duan, R.~Dwivedi, J.~Edwards, A.~Eirug, V.~Galanos, P.~V. Ilavarasan,
  M.~Janssen, P.~Jones, A.~K. Kar, H.~Kizgin, B.~Kronemann, B.~Lal, B.~Lucini,
  R.~Medaglia, K.~{Le Meunier-FitzHugh}, L.~C. {Le Meunier-FitzHugh}, S.~Misra,
  E.~Mogaji, S.~K. Sharma, J.~B. Singh, V.~Raghavan, R.~Raman, N.~P. Rana,
  S.~Samothrakis, J.~Spencer, K.~Tamilmani, A.~Tubadji, P.~Walton, and M.~D.
  Williams, ``Artificial intelligence ({AI}): Multidisciplinary perspectives on
  emerging challenges, opportunities, and agenda for research, practice and
  policy,'' {\em International Journal of Information Management}, vol.~57,
  p.~101994, 2021.

\bibitem{Kong2023}
Y.-X. Kong, R.-J. Wu, Y.-C. Zhang, and G.-Y. Shi, ``Utilizing statistical
  physics and machine learning to discover collective behavior on temporal
  social networks,'' {\em Information Processing \& Management}, vol.~60,
  no.~2, p.~103190, 2023.

\bibitem{Hornischer2022}
H.~Hornischer, P.~J. Pritz, J.~Pritz, M.~G. Mazza, and M.~Boos, ``Modeling of
  human group coordination,'' {\em Physical Review Research}, vol.~4, no.~2,
  p.~023037, 2022.

\bibitem{Sytnyk2021}
D.~Sytnyk and R.~Melnik, ``Mathematical models with nonlocal initial
  conditions: An exemplification from quantum mechanics,'' {\em Mathematical
  and Computational Applications}, vol.~26, no.~4, p.~73, 2021.

\bibitem{Chow2012}
S.-N. Chow, W.~Huang, Y.~Li, and H.~Zhou, ``Fokker–planck equations for a
  free energy functional or markov process on a graph,'' {\em Archive for
  Rational Mechanics and Analysis}, vol.~203, pp.~969 -- 1008, 2012.

\bibitem{Jarzynski2008}
C.~Jarzynski, ``Nonequilibrium work relations: foundations and applications,''
  {\em The European Physical Journal B}, vol.~64, pp.~331--340, 2008.

\bibitem{Conti2022}
D.~Conti and T.~Mora, ``Nonequilibrium dynamics of adaptation in sensory
  systems,'' {\em Physical Review E}, vol.~106, no.~5, p.~054404, 2022.

\bibitem{Roldan2015}
E.~Rold\'an, I.~Neri, M.~D\"orpinghaus, H.~Meyr, and F.~J\"ulicher, ``Decision
  making in the arrow of time,'' {\em Phys. Rev. Lett.}, vol.~115, p.~250602,
  Dec 2015.

\bibitem{Wu2002}
W.~Wu, M.~Black, Y.~Gao, M.~Serruya, A.~Shaikhouni, J.~Donoghue, and
  E.~Bienenstock, ``Neural decoding of cursor motion using a kalman filter,''
  {\em Advances in neural information processing systems}, vol.~15, 2002.

\bibitem{Vroylandt2022}
H.~Vroylandt, L.~Gouden{\`e}ge, P.~Monmarch{\'e}, F.~Pietrucci, and
  B.~Rotenberg, ``Likelihood-based non-markovian models from molecular
  dynamics,'' {\em Proceedings of the National Academy of Sciences}, vol.~119,
  no.~13, p.~e2117586119, 2022.

\bibitem{Stefan2022}
A.~M. Stefan, F.~D. Sch{\"o}nbrodt, N.~J. Evans, and E.-J. Wagenmakers,
  ``Efficiency in sequential testing: Comparing the sequential probability
  ratio test and the sequential bayes factor test,'' {\em Behavior Research
  Methods}, vol.~54, no.~6, pp.~3100--3117, 2022.

\bibitem{Roweis1999}
S.~Roweis and Z.~Ghahramani, ``A unifying review of linear gaussian models,''
  {\em Neural computation}, vol.~11, no.~2, pp.~305--345, 1999.

\bibitem{Bishop2007}
C.~M. Bishop and N.~M. Nasrabadi, {\em Pattern recognition and machine
  learning}, vol.~4.
\newblock Springer, 2007.

\bibitem{Gnesotto2020}
F.~S. Gnesotto, G.~Gradziuk, P.~Ronceray, and C.~P. Broedersz, ``Learning the
  non-equilibrium dynamics of brownian movies,'' {\em Nature communications},
  vol.~11, no.~1, p.~5378, 2020.

\bibitem{Seif2021}
A.~Seif, M.~Hafezi, and C.~Jarzynski, ``Machine learning the thermodynamic
  arrow of time,'' {\em Nature Physics}, vol.~17, pp.~105--113, 2021.

\bibitem{Perl2021}
Y.~S. Perl, H.~Bocaccio, C.~Pallavicini, I.~P{\'e}rez-Ipi{\~n}a, S.~Laureys,
  H.~Laufs, M.~L. Kringelbach, G.~Deco, and E.~Tagliazucchi, ``Nonequilibrium
  brain dynamics as a signature of consciousness.,'' {\em Physical review. E},
  vol.~104 1-1, p.~014411, 2021.

\bibitem{Pal2022}
S.~Pal and R.~Melnik, ``Nonlocal models in the analysis of brain
  neurodegenerative protein dynamics with application to alzheimer’s
  disease,'' {\em Scientific Reports}, vol.~12, no.~1, p.~7328, 2022.

\bibitem{Shaheen2022}
H.~Shaheen, S.~Pal, and R.~Melnik, ``Multiscale co-simulation of deep brain
  stimulation with brain networks in neurodegenerative disorders,'' {\em Brain
  Multiphysics}, vol.~3, p.~100058, 2022.

\bibitem{Pal2023}
S.~Pal and R.~Melnik, ``Non-markovian behaviour and the dual role of astrocytes
  in alzheimer's disease development and propagation,'' {\em arXiv preprint
  arXiv:2208.03540}, 2023.

\bibitem{Kensinger2003}
E.~A. Kensinger, D.~K. Shearer, J.~J. Locascio, J.~H. Growdon, and S.~Corkin,
  ``Working memory in mild alzheimer's disease and early parkinson's
  disease.,'' {\em Neuropsychology}, vol.~17, no.~2, p.~230, 2003.

\bibitem{Gagnon2011}
L.~G. Gagnon and S.~Belleville, ``Working memory in mild cognitive impairment
  and alzheimer's disease: contribution of forgetting and predictive value of
  complex span tasks.,'' {\em Neuropsychology}, vol.~25 2, pp.~226--236, 2011.

\bibitem{Stopford2012}
C.~L. Stopford, J.~C. Thompson, D.~Neary, A.~M. Richardson, and J.~S. Snowden,
  ``Working memory, attention, and executive function in alzheimer’s disease
  and frontotemporal dementia,'' {\em Cortex}, vol.~48, no.~4, pp.~429--446,
  2012.

\bibitem{Yan2020}
H.~Yan and J.~Wang, ``Non-equilibrium landscape and flux reveal the
  stability-flexibility-energy tradeoff in working memory,'' {\em PLoS
  Computational Biology}, vol.~16, no.~e1008209, 2020.

\bibitem{Murray2017}
J.~D. Murray, J.~Jaramillo, and X.-J. Wang, ``Working memory and
  decision-making in a frontoparietal circuit model,'' {\em Journal of
  Neuroscience}, vol.~37, no.~50, pp.~12167--12186, 2017.

\bibitem{Brent1978}
S.~B. Brent, ``Prigogine's model for self-organization in nonequilibrium
  systems: Its relevance for developmental psychology,'' {\em Human
  Development}, vol.~21, no.~5/6, pp.~374--387, 1978.

\bibitem{Chapman1991}
M.~Chapman, ``Self-organization as developmental process: Beyond the organismic
  and mechanistic models?,'' {\em Annals of theoretical psychology},
  pp.~335--348, 1991.

\bibitem{West2012}
B.~J. West and P.~Grigolini, ``Networking of psychophysics, psychology, and
  neurophysiology,'' 2012.

\bibitem{Boettcher2011}
S.~Boettcher and C.~Brunson, ``Renormalization group for critical phenomena in
  complex networks,'' {\em Frontiers in physiology}, vol.~2, p.~102, 2011.

\bibitem{Kaupuzs2022}
J.~Kaupužs and R.~V.~N. Melnik, ``Functional truncations for the solution of
  the nonperturbative rg equations,'' {\em Journal of Physics A: Mathematical
  and Theoretical}, vol.~55, no.~46, p.~465002, 2022.

\bibitem{Tadic2023}
B.~Tadi{\'c}, M.~M. Dankulov, and R.~Melnik, ``Evolving cycles and
  self-organised criticality in social dynamics,'' {\em Chaos, Solitons \&
  Fractals}, vol.~171, p.~113459, 2023.

\bibitem{Shizgal2018}
B.~D. Shizgal, ``Kappa and other nonequilibrium distributions from the
  fokker-planck equation and the relationship to tsallis entropy,'' {\em Phys.
  Rev. E}, vol.~97, p.~052144, 2018.

\bibitem{Tanaka2022}
S.~Tanaka, T.~Umegaki, A.~Nishiyama, and H.~Kitoh-Nishioka, ``Dynamical free
  energy based model for quantum decision making,'' {\em Physica A: Statistical
  Mechanics and its Applications}, vol.~605, p.~127979, 2022.

\bibitem{Johnson2017}
J.~Johnson, A.~Nowak, P.~Ormerod, B.~Rosewell, and Y.-C. Zhang, {\em
  Non-equilibrium social science and policy}.
\newblock Understanding Complex Systems book series, Springer, 2017.

\end{thebibliography}
\end{document}